# Robust concurrent topology optimization of structure and its composite material considering uncertainty with imprecise probability

Y. Wu[a], Eric Li[b], Z. C. He[a, c*], X. Y. Lin[a], H. X. Jiang[a]

[a] State Key Laboratory of Advanced Design and Manufacturing for Vehicle Body, Hunan University, Changsha, 410082, P. R. China

[b] School of Science, Engineering & Design, Teesside University, Middlesbrough, UK

[c] Guangxi Key Laboratory of Automobile Components and Vehicle technology, Guangxi University of Science and Technology, Liuzhou, 545006, P. R. China

## Abstract

This paper studied a robust concurrent topology optimization (RCTO) approach to design the structure and its composite materials simultaneously. For the first time, the material uncertainty with imprecise probability is integrated into the multi-scale concurrent topology optimization (CTO) framework. To describe the imprecise probabilistic uncertainty efficiently, the type I hybrid interval random model is adopted. An improved hybrid perturbation analysis (IHPA) method is formulated to estimate the expectation and stand variance of the objective function in the worst case. Combined with the bi-directional evolutionary structural optimization (BESO) framework, the robust designs of the structure and its composite material are carried out. Several 2D and 3D numerical examples are presented to illustrate the effectiveness of the proposed method. The results show that the proposed method has high efficiency and low precision loss. In addition, the proposed RCTO approach remains efficient in both of linear static and dynamic structures, which shows its extensive adaptability.

---

*Corresponding author: Tel./fax: +86 73188822051.

Email address: hezhicheng815@hnu.edu.cn (ZC He)

# 1. Introduction

Structural design and optimization have always played an important role in engineering. It is because how to find the optimal material distribution in a given design domain to get the best structural performance is the basic problem of structural design and optimization [1]. In 1988, Bendsøe and Kikuchi [2] first introduced a seminal method called topology optimization. Since then, the theory, method, and application of topology optimization have been developed rapidly [3-5].

Nowadays, with the development of computer technology and numerical method, topology optimization is no longer limited to the design of structures anymore. The distributions of the ideal materials on the micro scales have received a lot of attention [6, 7]. At the macro-scale, topology optimization focuses on achieving the best structural performance using predefined material. As for the micro-scale, topology optimization is usually adopted to seek the material distribution for prescribed or extreme effective properties [8]. The material design at the micro-scale assumes that the material is constructed by periodic unit cells (PUC), thus its effective properties can be calculated by the numerical homogenization theory [9]. The efficient material distribution on the micro-scale enriches the capacity of topology optimization method for more extensive advanced designs and applications [10, 11]. However, whether it is structural-oriented or material-oriented topology optimization, it is difficult to obtain optimal design of structure and material at the same time. To this end, an integrated topology optimization method called concurrent topology optimization (CTO) [12-16] was proposed, in which the optimal structure and material distribution can be designed simultaneously [17]. Inspired by the pioneer works [18, 19], many achievements have been made for different aspects of CTO. [20-24]. These studies show that CTO can evidently extend the design space than those using the single-scale based approaches, and further improve the structural performances, such as compliance [12], dynamic compliance [25], natural frequency [8, 26] and frequency responses [27].

Uncertainty is widespread in practical engineering [28], which has a significant effect on the prediction of structural performance. For example, a slight disturbance of Poisson's ratio of incompressible rubber has a great influence on the band gap of ternary acoustic metamaterials [29]. Generally, we can use probabilistic model to describe the uncertainty [30]. By using probability and statistics theories, a series of uncertainty analysis and design methods [31-33] have been well established. It is noted that the probabilistic models require a large number of experimental samples to establish accurate probability distributions of uncertain parameters, which puts high demands on practical applications. Due to the difficulty and high cost of testing in the extensive engineering, it is difficult to obtain the accurate probability distribution of the uncertain parameter [34], which means the imprecise probabilities are most likely acquired. Studies on uncertainty with imprecise probability can be dated back to 1980s [35], and a large variety of specific theories and mathematical models have been proposed [36]. The hybrid interval random that efficiently describes the imprecise probabilistic parameters, such as expectation and standard variance, by interval model [29, 37] can be considered as a feasible model for imprecise probability [38]. It is noted that there are two commonly known hybrid interval random models, in which the type I hybrid model can be adopted to deal with imprecise probability, while the type II hybrid model is generally employed for partially unmeasurable uncertainty [39]. The type I hybrid model not only improves the ability of the probabilistic based model to cope with complex uncertainties in a unified analysis, but also avoids the huge computational burden naturally [40], thus aroused great concerns in engineering analysis [41].

To further incorporate the uncertainty into the topology optimization, researchers have developed many methods. The reliability-based topology optimization (RBTO) [42-44] and the robust topology optimization (RTO) method [45-47] are widely used to deal with uncertainty effect. In the past few years, these two methods have been gradually introduced into CTO. Guo et al. [48] proposed a two-scale RTO for design of material and structure under unknown-but-bounded load uncertainties. Chan et al. [49] presented a RTO to design multiscale structure with probabilistic-based multi-material uncertainty. Deng et al. [50] investigated a RTO approach for the multiscale structure and multi porous with random field uncertainty. Wang et al. [51, 52] introduced two RBTO frameworks for the CTO of solid and truss-like structure considering unknown-but-bounded uncertainties. Besides, hybrid interval random model is also involved in the CTO. Zheng et al. [53, 54] developed two RTO approaches to find the robust hierarchical design of the structure with type II hybrid uncertainties, in which the probability distribution and variation interval are respectively employed to describe different parameters. It would be suitable for problems that partially have interval variational uncertain variable with unknown distribution type and parameter, but could not be efficient for imprecise probability [39]. Considering that the probability uncertainty which is difficult to be expressed by precise parameters in actual engineering, the CTO method considering uncertainty with imprecise probability is still to be developed.

To this end, a robust concurrent topology optimization (RCTO) considering uncertainty with imprecise probability is proposed for the first time in this paper. In our work, the macrostructure is composed of composite structures that fill two materials in micro-scale that naturally avoid the problem of continuity [55]. Based on the the type I hybrid interval random model, the imprecise probability of materials properties is formulated by probabilistic distribution with interval varied probabilistic parameters. To integrate the uncertainty into the CTO efficiently, an improved hybrid perturbation analysis (IHPA) method is formulated, and the worst case of the mean-compliance can be estimated. Meanwhile, the accuracy of the proposed method is disscussed dialectically. The uniform weight constraints are employed and compared with the designs obtained by guessing separate weight constraints. Besides, both of static and dynamic loaded structures are investigated to prove the applicability of the proposed RCTO approach in real world.

The following paragraphs are organized as follows: Section 2 briefly introduces the deterministic concurrent topology optimization (DCTO) method in bi-directional evolutionary structural optimization (BESO) framework. Section 3 descripts the uncertainty with imprecise probability and propose the improved hybrid perturbation analysis (IHPA) method based on the hybrid interval random model. Section 4 describes the RCTO approach in detail, including its formulations, sensitivity derivations and numerical implementation. Section 5 shows several numerical examples and Section 6 makes the final conclusions.

## 2. Deterministic concurrent topology optimization

### 2.1 Equilibrium equation

For the structures without damping under external excitation, the equilibrium equation can be expressed as

$$\mathbf{M}\ddot{\mathbf{U}}_t + \mathbf{K}\mathbf{U}_t = \mathbf{F}e^{j\omega_p t}, \tag{1}$$

where $\mathbf{M}$ and $\mathbf{K}$ denote the global mass and stiffness matrices respectively. $\mathbf{F}e^{j\omega_p t}$ represents the loading vector of the external excitation related to time $t$ and excitation frequency $\omega_p$. $\ddot{\mathbf{U}}_t$ and $\mathbf{U}_t$ are the structural accelerate and displacement vector caused by the external excitation. By assuming $\mathbf{U}_t = \mathbf{U}e^{j\omega_p t}$, where $\mathbf{U}$ denotes the amplitude of the displacement, Eq (1) can be rewritten as

$$\begin{cases} \mathbf{K}\mathbf{U} = \mathbf{F}, & \text{when } \omega_p = 0 \\ \left(\mathbf{K} - \omega_p^2 \mathbf{M}\right)\mathbf{U} = \mathbf{F}, & \text{when } \omega_p \neq 0 \end{cases} \tag{2}$$

Eq. (2) represents the equilibrium equation of two most basic situations in the practical engineering: the static loaded ($\omega_p = 0$) and harmonic excited ($\omega_p \neq 0$) structure. They are widespread in some engineering, such as automobile and aerospace. Considering that their equilibrium equations are very similar in linear cases, they will all be considered in this study.

### 2.2 Mathematical formulation

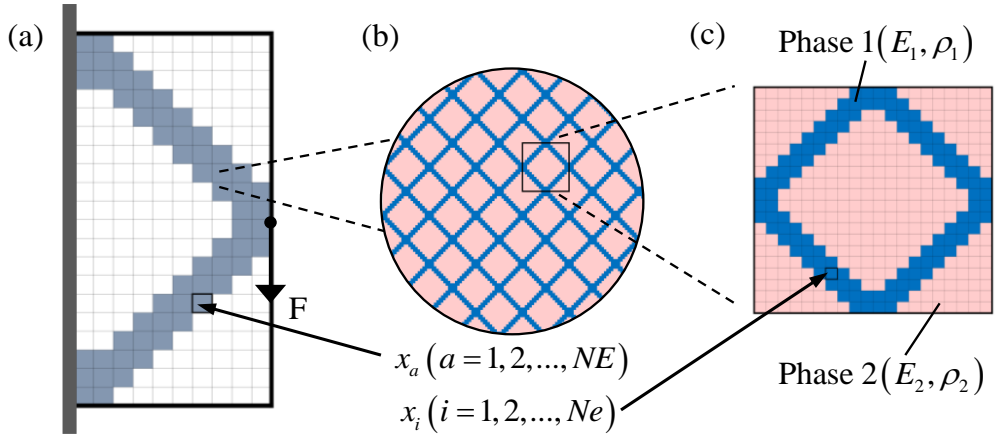

**Fig. 1.** Composite material composed two-scale structure: (a) macrostructure; (b) microstructure of composite material; (c) periodic unit cell (PUC).

Consider a two-scale structure as shown in Fig. 1 [56], it is assumed that the boundary condition and external excitation of the macrostructure are already known. Fig. 1(b) represents the microstructure of composite material that constructs the macrostructure. The microstructure is composed of two basic materials: phase 1 in blue and phase 2 in pink. Assume that the composite material is constituted by periodic unit cell (PUC), which can be represented by Fig. 1(c). $x_a$, $x_i$ and *NE*, *Ne* denote the design variables and their amount at macro-scale and micro-scale, respectively. For the convenience of distinction, we assume the density and elasticity modulus of two phases are $\rho_1$, $E_1$ and $\rho_2$, $E_2$ respectively ($\rho_1 > \rho_2, E_1 > E_2$).

The DCTO for two-scale structure can be formulated mathematically as follows

$$\text{Find: } x_a, x_i \left( a = 1, 2, ..., NE; \ i = 1, 2, ..., Ne \right)$$

$$\text{Minimize: } C = \mathbf{F}^{\mathrm{T}} \mathbf{U}$$

$$\text{Subject to: } \begin{cases} \mathbf{KU} = \mathbf{F}, & \text{when } \omega_p = 0, \\ \left( \mathbf{K} - \omega_p^2 \mathbf{M} \right) \mathbf{U} = \mathbf{F}, & \text{when } \omega_p \neq 0, \end{cases} \tag{3}$$

$$m\left( x_a, x_i \right) - W_f^* m_0 = 0,$$

$$\text{where: } x_a, x_i = x_{\min} \text{ or } 1 \left( 0 < x_{\min} \le x_a \le 1 \right).$$

where the objective function $C$ is the structural mean-compliance. $W_f^*$ denotes the target weight fraction of the design domain. $m_0 = \sum_{a=1}^{NE} V_a \rho_1$ represents the initial phase 1 full filled design, where $V_a$ denotes the volume of $a$-th element. $m$ is the designed total weight, which can be expressed by

$$m = \sum_{a=1}^{NE} x_a V_a \rho^H \left( x_i \right), \tag{4}$$

where $\rho^H \left( x_i \right)$ will be defined in Eq. (6a).

At the micro-scale, $x_i = x_{\min}$ or $1$ identifies the distribution of the two-phase composite, in which $x_i = 1$ defines the $i$-th element as phase 1, otherwise it is filled by phase 2. By adopting the solid isotropic material with penalization (SIMP) [3] scheme, the density and elasticity matrix of the $i$-th element can be associated with design variable $x_i$ as

$$\rho_i \left( x_i \right) = x_i \rho_1 + \left( 1 - x_i \right) \rho_2, \tag{5a}$$

$$\mathbf{D}_i \left( x_i \right) = x_i^p \mathbf{D}_1 + \left( 1 - x_i^p \right) \mathbf{D}_2, \tag{5b}$$

where $\rho_1$, $\rho_2$ and $\mathbf{D}_1$, $\mathbf{D}_2$ denote the density and elasticity matrix of phase 1 and phase 2 respectively. $p$ is the penalization index. For each PUC, their effective density and elasticity matrix $\rho^H \left( x_i \right)$ and $\mathbf{D}^H \left( x_i \right)$ can be obtained from the numerical homogenization theory [9] when the base cell is very small compared to the size of the structure. The effective properties of the PUC with material interpolation scheme are directly given by

$$\rho^H(x_i) = \frac{1}{|Y|} \sum_{i=1}^{Ne} V_i \left[ x_i \rho_1 + (1-x_i) \rho_2 \right] \tag{6a}$$

$$\mathbf{D}^H(x_i) = \frac{1}{|Y|} \sum_{i=1}^{Ne} \int_Y (\boldsymbol{\varepsilon}_0 - \boldsymbol{\varepsilon})^{\mathrm{T}} \left[ x_i^p \mathbf{D}_1 + (1-x_i^p) \mathbf{D}_2 \right] (\boldsymbol{\varepsilon}_0 - \boldsymbol{\varepsilon}) dY , \tag{6b}$$

where $V_i$ is the volume of the $i$-th element on micro-scale. $|Y|$ denotes the total area (for 2D cases) or volume (for 3D cases) of the PUC. $\boldsymbol{\varepsilon}_0$ are three unit test strains, e.g. $\{1, 0, 0\}^{\mathrm{T}}$, $\{0, 1, 0\}^{\mathrm{T}}$ $\{0, 0, 1\}^{\mathrm{T}}$ for 2D case. The strain fields $\boldsymbol{\varepsilon}$ are induced by these test strains under the periodical boundary conditions. It can be seen that the effective density and elasticity matrix depend on the distribution of design variable $x_i$. To get the detailed derivation process of Eq. (6), readers may refer to [8, 57, 58].

At the macro-scale, $x_a = x_{\min}$ or 1 determines whether the $a$-th element is void or solid. $x_a = 1$ determines the $a$-th as solid element, which is constructed by two-phase composite. Otherwise, it is the void element. Considering that a small value of $x_a$ will result extremely high ratio between penalization on mass and stiffness [59], which leads to the artificial local mode phenomenon in the low-density area. An alternative interpolation scheme [60] is adopted to maintain the ratio between mass and stiffness as follows

$$\rho_a(x_a, x_i) = x_a \rho^H(x_i), \tag{7a}$$

$$\mathbf{D}_a(x_a, x_i) = \left[ \frac{x_{\min} - x_{\min}^p}{1 - x_{\min}^p} (1 - x_a^p) + x_a^p \right] \mathbf{D}^H(x_i), \tag{7b}$$

where $\rho_a(x_a, x_i)$ and $\mathbf{D}_a(x_a, x_i)$ are the density and elasticity matrix of the $a$-th element.

### 2.3 Sensitivity analysis and sensitivity number

At the macro-scale, the elemental sensitivity of structural mean-compliance with respect to design variable $x_a$ can be derived by the adjoint method [61] as

$$\frac{\partial C}{\partial x_a} = -\mathbf{U}_a^{\mathrm{T}} \left( \frac{\partial \mathbf{K}_a}{\partial x_a} - \omega^2 \frac{\partial \mathbf{M}_a}{\partial x_a} \right) \mathbf{U}_a , \tag{8}$$

where $\mathbf{K}_a$, $\mathbf{M}_a$ and $\mathbf{U}_a$ denote the stiffness matrix, mass matrix and nodal displacement vector of the $a$-th element on the macro-scale, in which $\mathbf{K}_a$ and $\mathbf{M}_a$ can be represented by

$$\mathbf{K}_a = \int_A \mathbf{B}^{\mathrm{T}} \mathbf{D}_a\left(x_a, x_i\right) \mathbf{B} dA , \tag{9a}$$

$$\mathbf{M}_a = \int_A \rho_a\left(x_a, x_i\right) \mathbf{N}^{\mathrm{T}} \mathbf{N} dA , \tag{9b}$$

where $\mathbf{B}$ denotes the strain-displacement matrix. $\mathbf{N}$ represents the elemental shape function matrix. The symbol $A$ expresses the area or volume of related element. The derivations $\dfrac{\partial \mathbf{K}_a}{\partial x_a}$ and $\dfrac{\partial \mathbf{M}_a}{\partial x_a}$ can be directly obtained from Eqs. (7) and (9) as

$$\frac{\partial \mathbf{K}_a}{\partial x_a} = p x_a^{p-1} \int_A \mathbf{B}^{\mathrm{T}} \mathbf{D}^H\left(x_i\right) \mathbf{B} dA \tag{10a}$$

$$\frac{\partial \mathbf{M}_a}{\partial x_a} = \int_A \rho^H\left(x_i\right) \mathbf{N}^{\mathrm{T}} \mathbf{N} dA , \tag{10b}$$

Similarly, the elemental sensitivity at the micro-scale with respect to design variable $x_i$ can be derived as

$$\frac{\partial C}{\partial x_i} = -\sum_{a=1}^{NE} \mathbf{U}_a^{\mathrm{T}} \left( \frac{\partial \mathbf{K}_a}{\partial x_i} - \omega^2 \frac{\partial \mathbf{M}_a}{\partial x_i} \right) \mathbf{U}_a , \tag{11}$$

where $\dfrac{\partial \mathbf{K}_a}{\partial x_i}$ and $\dfrac{\partial \mathbf{M}_a}{\partial x_i}$ can be derived from Eqs. (6-7) and (9) as

$$\frac{\partial \mathbf{K}_a}{\partial x_i} = \left( \frac{x_{\min} - x_{\min}^p}{1 - x_{\min}^p}\left(1 - x_a^p\right) + x_a^p \right) \int_A \mathbf{B}^T \left( \frac{p x_i^{p-1}}{|Y|} \int_Y (\boldsymbol{\varepsilon}_0 - \boldsymbol{\varepsilon})^{\mathrm{T}} (\mathbf{D}_1 - \mathbf{D}_2)(\boldsymbol{\varepsilon}_0 - \boldsymbol{\varepsilon}) dY \right) \mathbf{B} dA , \tag{12a}$$

$$\frac{\partial \mathbf{M}_a}{\partial x_i} = x_a \int_A \left( \frac{1}{|Y|} \sum_{i=1}^{Ne} V_i (\rho_1 - \rho_2) \right) \mathbf{N}^{\mathrm{T}} \mathbf{N} dA , \tag{12b}$$

In the ESO/BESO method, the structure is optimized using discrete design variables. The discrete sensitivity number can be obtained by processing the results of the sensitivity analysis. At the macro-scale, the elemental sensitivity number of the discrete design variable $x_a = x_{\min}$ or 1 can be determined by

$$\alpha_a = -\frac{1}{p} \frac{\partial C}{\partial x_a} = \begin{cases} \mathbf{U}_a^{\mathrm{T}} \left( \int_A \mathbf{B}^T \mathbf{D}^H \mathbf{B} dA - \omega^2 \int_A \rho^H \mathbf{N}^{\mathrm{T}} \mathbf{N} dA \right) \mathbf{U}_a , & \text{when } x_a = 1 \\ \mathbf{U}_a^{\mathrm{T}} \left( x_{\min}^{p-1} \int_A \mathbf{B}^T \mathbf{D}^H \mathbf{B} dA - \omega^2 \int_A \rho^H \mathbf{N}^{\mathrm{T}} \mathbf{N} dA \right) \mathbf{U}_a , & \text{when } x_a = x_{\min} \end{cases} \tag{13}$$

At the micro-scale, the microstructure is constructed by two-phase material, thus the elemental sensitivity number of the discrete design variable $x_i = x_{\min}$ or $1$ can be calculated by

$$
\alpha_i = -\frac{1}{p}\frac{\partial C}{\partial x_i} =
\begin{cases}
\sum_{a=1}^{NE}\mathbf{U}_a^{\mathrm{T}}\left(\left(\dfrac{x_{\min}-x_{\min}^p}{1-x_{\min}^p}\left(1-x_a^p\right)+x_a^p\right)\int_A \mathbf{B}^T\left(\dfrac{1}{|Y|}\int_Y\left(\boldsymbol{\varepsilon}_0-\boldsymbol{\varepsilon}\right)^{\mathrm{T}}\left(\mathbf{D}_1-\mathbf{D}_2\right)\left(\boldsymbol{\varepsilon}_0-\boldsymbol{\varepsilon}\right)dY\right)\mathbf{B}dA\right. \\
\qquad\qquad \left.-\omega^2\dfrac{x_a}{p}\left(\dfrac{1}{|Y|}\sum_{i=1}^{Ne}V_i\left(\rho_1-\rho_2\right)\right)\int_A\mathbf{N}^{\mathrm{T}}\mathbf{N}dA\right)\mathbf{U}_a, & \text{when } x_i=1 \\[2em]
\sum_{a=1}^{NE}\mathbf{U}_a^{\mathrm{T}}\left(\left(\dfrac{x_{\min}-x_{\min}^p}{1-x_{\min}^p}\left(1-x_a^p\right)+x_a^p\right)\int_A \mathbf{B}^T\left(\dfrac{x_{\min}^{p-1}}{|Y|}\int_Y\left(\boldsymbol{\varepsilon}_0-\boldsymbol{\varepsilon}\right)^{\mathrm{T}}\left(\mathbf{D}_1-\mathbf{D}_2\right)\left(\boldsymbol{\varepsilon}_0-\boldsymbol{\varepsilon}\right)dY\right)\mathbf{B}dA\right. \\
\qquad\qquad \left.-\omega^2\dfrac{x_a}{p}\left(\dfrac{1}{|Y|}\sum_{i=1}^{Ne}V_i\left(\rho_1-\rho_2\right)\right)\int_A\mathbf{N}^{\mathrm{T}}\mathbf{N}dA\right)\mathbf{U}_a, & \text{when } x_i=x_{\min}
\end{cases}
\tag{14}
$$

## 3. Treatment of imprecise probability

### 3.1. Description of imprecise probability

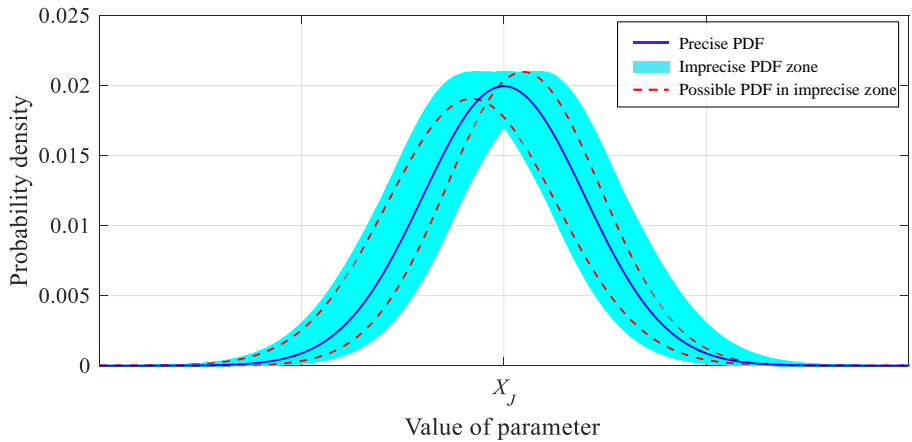

Fig. 2 The probability distribution functions of precise and imprecise probability with normal distribution.

For the probabilistic-based uncertainty, the probability distribution function (PDF) can be adopted to describe the uncertain parameters. However, in practical engineering, the precise PDF can hardly be obtained due to the lack or poor quality of information [39]. As shown in Fig. 2, the uncertainty parameter $X_J$ is assumed to be in normal distribution but lack of information. Thus, the possible PDFs are shown, which constitute a zone in cyan. Two of the possible PDFs are presented by red dash line. Compared with the PDF with accurate distribution parameter colored in blue, it is obvious that the precise probability distribution function has limitations and cannot reflect all the situations.

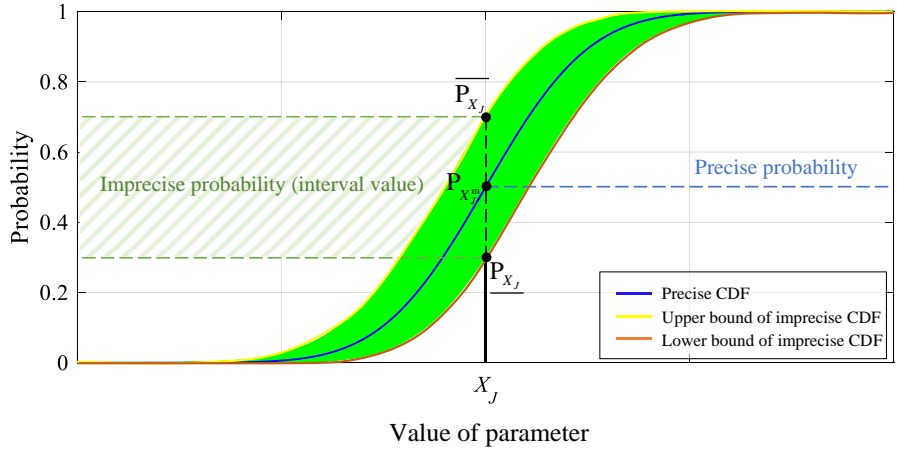

Fig. 3 P-box model for the description of uncertainty with imprecise probability.

To further describe the imprecise probabilistic uncertainty, a p-box [62] model is plotted by adopting the information in Fig. 2. The corresponding cumulative distribution functions (CDF) are shown in Fig. 3 in green. It can be seen that the probability of imprecise uncertainty is located within the 'strip', which can be represented by an interval. By determining the upper and lower bound of the imprecise uncertainty, the probability range of the parameters is clarified. For an arbitrary parameter $X_J$, its probability can be expressed by $\left[ \overline{\mathrm{P}_{X_J}}, \underline{\mathrm{P}_{X_J}} \right]$. Fig. 3 indicates that the interval model has potential to be integrated into the probabilistic-based model to describe the imprecise probability.

### 3.2. Hybrid interval random model

Assume that all the uncertain parameters are independent. $X$ represents an uncertain probabilistic parameter that has imprecise probability. By adopting the interval model to describe $X$, $X(\mathbf{Y})$ can be obtained. Without loss of generality, let $X_J(\mathbf{Y}_J)$ be the $J$-th hybrid interval random variable of the hybrid interval random vector $\mathbf{X}(\mathbf{Y})$ that composed of all independent hybrid interval random variables. $\mathbf{X}(\mathbf{Y})$ can be formulated mathematically as below

$$\mathbf{X}(\mathbf{Y}) = \left( X_1(\mathbf{Y}_1), X_2(\mathbf{Y}_2), \ldots, X_J(\mathbf{Y}_J), \ldots \right), \quad J = 1, 2, \ldots, M , \tag{15a}$$

$$\mathbf{Y}_J = \left( Y_J^1, Y_J^2, \ldots, Y_J^K, \ldots \right), \quad K = 1, 2, \ldots, N , \tag{15b}$$

where $M$ and $N$ represent the number of random and interval parameters, respectively. For each interval vector, $\mathbf{Y}_J$ can be expressed as follows

$$\mathbf{Y}_J = \left[ \underline{\mathbf{Y}_J}, \overline{\mathbf{Y}_J} \right] = Y_J^{\mathrm{m}} + \mathbf{Y}_J^{\mathrm{I}}, \tag{16a}$$

$$Y_J^{\mathrm{m}} = \frac{\underline{\mathbf{Y}_J} + \overline{\mathbf{Y}_J}}{2}, \tag{16b}$$

$$\mathbf{Y}_J^{\mathrm{I}} = \left[ -\Delta Y_J, +\Delta Y_J \right], \tag{16c}$$

$$\Delta Y_J = \frac{\overline{\mathbf{Y}_J} - \underline{\mathbf{Y}_J}}{2}, \tag{16d}$$

where $\underline{\mathbf{Y}_J}$ and $\overline{\mathbf{Y}_J}$ denote the lower and upper bounds of interval vector $\mathbf{Y}_J$. The symbol $Y_J^{\mathrm{m}}$ is the mean value of $\mathbf{Y}_J$, which can be calculated by averaging the lower and upper bounds value as shown in Eq. (16b). $\mathbf{Y}_J^{\mathrm{I}}$ denotes the variation interval of $\mathbf{Y}_J$, which depends on the difference of the lower and upper bound values as shown in Eq. (16c). The deviation $\Delta Y_J$ of the symmetrical interval can be acquired by averaging the upper and lower bounds of $\mathbf{Y}_J$ as shown in Eq. (16d).

In the combined form, the $J$-th hybrid interval random parameter $X_J(\mathbf{Y}_J)$ can be formulated as

$$X_J(\mathbf{Y}_J) = X_J(Y_J^{\mathrm{m}}) + X_J(\mathbf{Y}_J^{\mathrm{I}}) \tag{17a}$$

$$X_J(Y_J^{\mathrm{m}}) = \frac{X_J(\underline{\mathbf{Y}_J}) + X_J(\overline{\mathbf{Y}_J})}{2}, \tag{17b}$$

$$X_J(\mathbf{Y}_J^{\mathrm{I}}) = \left[ -\Delta X_J(\mathbf{Y}_J), +\Delta X_J(\mathbf{Y}_J) \right], \tag{17c}$$

$$\Delta X_J(\mathbf{Y}_J) = \frac{X_J(\overline{\mathbf{Y}_J}) - X_J(\underline{\mathbf{Y}_J})}{2}. \tag{17d}$$

The expectation and standard variance of the $J$-th hybrid interval random parameter $X_J(\mathbf{Y}_J)$ can be expressed as $\mu(X_J(\mathbf{Y}_J))$ and $\sigma(X_J(\mathbf{Y}_J))$ respectively. By adopting Eq. (17), the interval expression of $\mu(X_J(\mathbf{Y}_J))$ and $\sigma(X_J(\mathbf{Y}_J))$ can be represented.

### 3.3. Improved hybrid perturbation analysis (IHPA) method

By assuming the deterministic loading, the equilibrium equation in Eq. (2) with material uncertainty can be represented as

$$\begin{cases} \mathbf{K}(\mathbf{X}(\mathbf{Y}))\mathbf{U}(\mathbf{X}(\mathbf{Y})) = \mathbf{F}, & \text{when } \omega_p = 0 \\ \left(\mathbf{K}(\mathbf{X}(\mathbf{Y})) - \omega_p^2 \mathbf{M}(\mathbf{X}(\mathbf{Y}))\right)\mathbf{U}(\mathbf{X}(\mathbf{Y})) = \mathbf{F}, & \text{when } \omega_p \neq 0 \end{cases} \tag{18}$$

For simplicity, we use $\mathbf{K}_d(\mathbf{X}(\mathbf{Y}))$ to represent $\mathbf{K}(\mathbf{X}(\mathbf{Y}))$ and $\mathbf{K}(\mathbf{X}(\mathbf{Y})) - \omega_p^2 \mathbf{M}(\mathbf{X}(\mathbf{Y}))$ in Eq. (18). The structural mean-compliance with imprecise probability can be represented as below

$$C(\mathbf{X}(\mathbf{Y})) = \mathbf{F}^{\mathrm{T}} \mathbf{U}(\mathbf{X}(\mathbf{Y})) \tag{19}$$

where $\mathbf{U}(\mathbf{X}(\mathbf{Y}))$ is caused by the uncertainty.

In the IHPA method, we firstly assume the interval variables related to $\mathbf{X}(\mathbf{Y})$ are deterministic. The first-order Taylor series expansion of $\mathbf{U}(\mathbf{X}(\mathbf{Y}))$ at the expectation of the interval random parameter vector $\mathbf{X}(\mathbf{Y})$ can be expressed as

$$\mathbf{U}(\mathbf{X}(\mathbf{Y})) = \mathbf{U}(\mu(\mathbf{X}(\mathbf{Y}))) + \sum_{J=1}^{M} \frac{\partial \mathbf{U}(\mathbf{X}(\mathbf{Y}))}{\partial X_J(\mathbf{Y}_J)} \bigg|_{X_J(\mathbf{Y}_J) = \mu(X_J(\mathbf{Y}_J))} \left(X_J(\mathbf{Y}_J) - \mu(X_J(\mathbf{Y}_J))\right) + \mathrm{o}(X_J(\mathbf{Y}_J)). \tag{20}$$

As the variation of parameter is relatively small to itself, the remainder of the first-order Taylor series expansion $\mathrm{o}(X_J(\mathbf{Y}_J))$ in Eq. (20) can be ignored. Based on the random moment method [63], Eq. (20) can be expressed by two parts as

$$E(\mathbf{U}) = \mathbf{U}(\mu(\mathbf{X}(\mathbf{Y}))), \tag{21a}$$

$$SD(\mathbf{U}) = \sum_{J=1}^{M} \frac{\partial \mathbf{U}(\mathbf{X}(\mathbf{Y}))}{\partial X_J(\mathbf{Y}_J)} \bigg|_{X_J(\mathbf{Y}_J) = \mu(X_J(\mathbf{Y}_J))} \sigma(X_J(\mathbf{Y}_J)), \tag{21b}$$

where $E(\mathbf{U})$ and $SD(\mathbf{U})$ denote the expectation and standard variance, respectively. $\sigma(X_J(\mathbf{Y}_J))$ is equal to $X_J(\mathbf{Y}_J) - \mu(X_J(\mathbf{Y}_J))$ in Eq. (20).

As the interval variables are considered, both expectation $E(\mathbf{U})$ and standard variance $SD(\mathbf{U})$ are the interval vectors. Performing the first-order Taylor series expansion again yields

$$E(\mathbf{U}) = \mathbf{U}\Big(\mu\big(\mathbf{X}(\mathbf{Y}^{\mathrm{m}})\big)\Big) + \sum_{J=1}^{M}\sum_{K=1}^{N} \frac{\partial \mathbf{U}\big(\mu\big(X_J(\mathbf{Y}_J)\big)\big)}{\partial \mu\big(X_J(\mathbf{Y}_J^1)\big)}\Bigg|_{\mu\big(X_J(\mathbf{Y}_J^1)\big)=\mu\big(X_J(Y_J^{\mathrm{m}})\big)} \Big(\mu\big(X_J(Y_J^K)\big) - \mu\big(X_J(Y_J^{\mathrm{m}})\big)\Big) + \mathrm{o}\Big(\mu\big(X_J(\mathbf{Y}_J^1)\big)\Big)$$

$$(22a)$$

$$\begin{aligned}
SD(\mathbf{U}) = \sum_{J=1}^{M}\Bigg\{ &\frac{\partial \mathbf{U}\big(X_J(\mathbf{Y}_J)\big)}{\partial X_J(\mathbf{Y}_J)}\Bigg|_{X_J(\mathbf{Y}_J)=\mu\big(X_J(\mathbf{Y}_J)\big)} \sigma\big(X_J(Y_J^{\mathrm{m}})\big) \\
&+ \sum_{K=1}^{N}\Bigg[ \frac{\partial^2 \mathbf{U}\big(X_J(\mathbf{Y}_J)\big)}{\partial X_J(\mathbf{Y}_J)\,\partial X_J(\mathbf{Y}_J^1)}\Bigg|_{\substack{X_J(\mathbf{Y}_J)=\mu(X_J(\mathbf{Y}_J))\\ X_J(\mathbf{Y}_J^1)=X_J(Y_J^{\mathrm{m}})}} \big(\mu\big(X_J(Y_J^{\mathrm{m}})\big)-\mu\big(X_J(Y_J^K)\big)\big)\sigma\big(X_J(Y_J^{\mathrm{m}})\big) \\
&+ \frac{\partial \mathbf{U}\big(X_J(\mathbf{Y}_J)\big)}{\partial X_J(\mathbf{Y}_J)}\Bigg|_{X_J(\mathbf{Y}_J)=\mu(X_J(\mathbf{Y}_J))} \frac{\partial \sigma\big(X_J(Y_J^{\mathrm{m}})\big)}{\partial X_J(\mathbf{Y}_J^1)}\big(\mu\big(X_J(Y_J^K)\big)-\mu\big(X_J(Y_J^{\mathrm{m}})\big)\big)\Bigg] + \mathrm{o}\big(X_J(\mathbf{Y}_J^1)\big)\Bigg\}
\end{aligned}$$

$$(22b)$$

where the remainders $\mathrm{o}\big(\mu\big(X_J(\mathbf{Y}_J^1)\big)\big)$ and $\mathrm{o}\big(X_J(\mathbf{Y}_J^1)\big)$ can be ignored as the variation of parameter is relatively small to itself. For simplicity, Eq. (22) can be rewritten as

$$E(\mathbf{U}) \approx \mathbf{U}_0 + \sum_{J=1}^{M}\sum_{K=1}^{N} \mathbf{U}_{1,J}\Delta\mu\big(X_J(Y_J^K)\big) \tag{23a}$$

$$SD(\mathbf{U}) \approx \sum_{J=1}^{M}\Bigg(\mathbf{U}_{2,J}\sigma\big(X_J(Y_J^{\mathrm{m}})\big) + \sum_{K=1}^{N}\Big(\mathbf{U}_{3,JK}\sigma\big(X_J(Y_J^{\mathrm{m}})\big)\Delta\mu\big(X_J(Y_J^K)\big) + \mathbf{U}_{2,J}\Delta\sigma\big(X_J(Y_J^K)\big)\Big)\Bigg) \tag{23b}$$

where

$$\mathbf{U}_0 = \mathbf{U}\Big(\mu\big(\mathbf{X}(\mathbf{Y}^{\mathrm{m}})\big)\Big), \tag{24a}$$

$$\mathbf{U}_{1,J} = \frac{\partial \mathbf{U}\big(\mu\big(X_J(\mathbf{Y}_J)\big)\big)}{\partial \mu\big(X_J(\mathbf{Y}_J^1)\big)}\Bigg|_{\mu\big(X_J(\mathbf{Y}_J^1)\big)=\mu\big(X_J(Y_J^{\mathrm{m}})\big)}, \tag{24b}$$

$$\mathbf{U}_{2,J} = \frac{\partial \mathbf{U}\big(X_J(\mathbf{Y}_J)\big)}{\partial X_J(\mathbf{Y}_J)}\Bigg|_{X_J(\mathbf{Y}_J)=\mu\big(X_J(\mathbf{Y}_J)\big)}, \tag{24c}$$

$$\mathbf{U}_{3,JK} = \frac{\partial^2 \mathbf{U}\big(X_J(\mathbf{Y}_J)\big)}{\partial X_J(\mathbf{Y}_J)\,\partial X_J(\mathbf{Y}_J^1)}\Bigg|_{\substack{X_J(\mathbf{Y}_J)=\mu(X_J(\mathbf{Y}_J))\\ X_J(\mathbf{Y}_J^1)=X_J(Y_J^{\mathrm{m}})}}, \tag{24d}$$

and

$$\Delta\mu\left(X_J\left(Y_J^K\right)\right) = \frac{\partial X_J\left(Y_J^1\right)}{\partial\mu\left(X_J\left(Y_J^1\right)\right)}\left(\mu\left(X_J\left(Y_J^m\right)\right) - \mu\left(X_J\left(Y_J^K\right)\right)\right), \tag{25a}$$

$$\Delta\sigma\left(X_J\left(Y_J^K\right)\right) = \frac{\partial\sigma\left(X_J\left(Y_J^m\right)\right)}{\partial X_J\left(Y_J^K\right)}\left(\mu\left(X_J\left(Y_J^m\right)\right) - \mu\left(X_J\left(Y_J^K\right)\right)\right). \tag{25b}$$

where $\mathbf{U}_0$ and its derivations $\mathbf{U}_{1,J}$, $\mathbf{U}_{2,J}$ and $\mathbf{U}_{3,JK}$ can be calculated by substituting the mean value of expectation of the imprecise probability $\mu\left(\mathbf{X}\left(\mathbf{Y}^m\right)\right)$ into Eq. (18) as

$$\mathbf{U}_0 = \mathbf{K}_d^{-1}\left(\mu\left(\mathbf{X}\left(\mathbf{Y}^m\right)\right)\right)\mathbf{F}, \tag{26a}$$

$$\mathbf{U}_{1,J} = -\mathbf{K}_d^{-1}\left(\mu\left(\mathbf{X}\left(\mathbf{Y}^m\right)\right)\right)\frac{\partial\mathbf{K}_d\left(\mu\left(X_J\left(\mathbf{Y}_J\right)\right)\right)}{\partial\mu\left(X_J\left(Y_J^1\right)\right)}\mathbf{U}_0, \tag{26b}$$

$$\mathbf{U}_{2,J} = -\mathbf{K}_d^{-1}\left(\mu\left(\mathbf{X}\left(\mathbf{Y}^m\right)\right)\right)\frac{\partial\mathbf{K}_d\left(X_J\left(\mathbf{Y}_J\right)\right)}{\partial X_J\left(\mathbf{Y}_J\right)}\mathbf{U}_0, \tag{26c}$$

$$\mathbf{U}_{3,JK} = -\mathbf{K}_d^{-1}\left(\mu\left(\mathbf{X}\left(\mathbf{Y}^m\right)\right)\right)\left(2\frac{\partial\mathbf{K}_d\left(\mu\left(X_J\left(\mathbf{Y}_J\right)\right)\right)}{\partial X_J\left(\mathbf{Y}_J\right)}\mathbf{U}_{2,J} + \frac{\partial^2\mathbf{K}_d\left(\mu\left(X_J\left(\mathbf{Y}_J\right)\right)\right)}{\partial X_J\left(\mathbf{Y}_J\right)\partial X_J\left(Y_J^1\right)}\mathbf{U}_0\right). \tag{26d}$$

According to Eqs. (19) and (23), the expectation and standard variance of the maximal mean-compliance can be expressed by

$$\overline{E}(C) = \mathbf{F}^T\left(\mathbf{U}_0 + \sum_{J=1}^M\sum_{K=1}^N\mathbf{U}_{1,J}\Delta\mu\left(X_J\left(Y_J^K\right)\right)S_{1,JK}\right), \tag{27a}$$

$$\overline{SD}(C) = \mathbf{F}^T\left(\sum_{J=1}^M\left(\mathbf{U}_{2,J}\sigma\left(X_J\left(Y_J^m\right)\right)\right)S_{2,J} + \sum_{K=1}^N\left(\mathbf{U}_{3,JK}\sigma\left(X_J\left(Y_J^m\right)\right)\Delta\mu\left(X_J\left(Y_J^K\right)\right) + \mathbf{U}_{2,J}\Delta\sigma\left(X_J\left(Y_J^K\right)\right)\right)S_{3,JK}\right) \tag{27b}$$

where $S_{1,JK}$, $S_{2,J}$ and $S_{3,JK}$ represent the sign of relative part respectively

$$S_{1,JK} = sign\left(\mathbf{U}_{1,J}\Delta\mu\left(X_J\left(Y_J^K\right)\right)\right)$$

$$S_{2,J} = sign\left(\mathbf{U}_{2,J}\sigma\left(X_J\left(Y_J^m\right)\right)\right) \tag{28}$$

$$S_{3,JK} = sign\left(\mathbf{U}_{3,JK}\sigma\left(X_J\left(Y_J^m\right)\right)\Delta\mu\left(X_J\left(Y_J^K\right)\right) + \mathbf{U}_{2,J}\Delta\sigma\left(X_J\left(Y_J^K\right)\right)\right).$$

By adopting the IHPA method, the mean-compliance in the worst case (objective function in this paper) can be quickly evaluated. It is necessary to make a further discussion on the IHPA method. On the one hand, due to the high efficiency of the first-order Taylor series expansion, the number of finite element analysis (FEA) requests for the robust objective function calculated by IHPA is greatly reduced. For a problem with $n$ hybrid interval random variables, the total number of FEA calls is

$$FEAcalls = 1 + 3n. \tag{29}$$

The amount of FEA calls increases linearly with the increase of the variable. At the same amount of FEA calls, its accuracy is much higher than that of the Monte Carlo simulation (MCS) based method. In view of the high computational cost of CTO, it is very suitable for the uncertainty involved topology optimization. On the other hand, the first-order Taylor series expansion based perturbation method has been approved to be accurate for linear uncertainty with small variation ranges [41, 64]. Generally speaking, the IHPA method makes CTO possible to take imprecise probabilistic parameters into account, and at the same time, to ensure the accuracy for evaluating uncertainties with small variation at low computational cost. The computational efficiency of this method will be demonstrated in the first numerical example, in which IHPA is compared with the MCS.

## 4. Robust concurrent topology optimization

### 4.1 Mathematical formulation

Based on the IHPA method, the robust concurrent topology optimization (RCTO) approach can be mathematically stated as

$$
\begin{aligned}
&\text{Find: } x_a, x_i \left( a = 1, 2, ..., NE; \ i = 1, 2, ..., Ne \right) \\
&\text{Minimize: } \overline{C} = \overline{E} \left( C \left( \mathbf{X}(\mathbf{Y}) \right) \right) + \kappa \overline{SD} \left( C \left( \mathbf{X}(\mathbf{Y}) \right) \right) \\
&\text{Subject to: } C \left( \mathbf{X}(\mathbf{Y}) \right) = \mathbf{F}^{\mathrm{T}} \mathbf{U} \left( \mathbf{X}(\mathbf{Y}) \right), \\
&\qquad\qquad \mathbf{K}_{\mathrm{d}} \left( \mathbf{X}(\mathbf{Y}) \right) \mathbf{U} \left( \mathbf{X}(\mathbf{Y}) \right) = \mathbf{F}, \\
&\qquad\qquad m \left( x_a, x_i \right) - W_f^* m_0 = 0, \\
&\text{where: } x_a, x_i = x_{\min} \text{ or } 1 \left( 0 < x_{\min} \leq x_a, x_i \leq 1 \right),
\end{aligned}
\tag{30}
$$

where $\overline{C}$ denotes the objective function, which is composed by maximal expectation and weighted standard variance. $\kappa$ is the weighting parameter, which is also called the robust optimization parameter. The rest of symbols in Eq. (30) have been introduced in the previous text.

## 4.2 Sensitivity number

The sensitivity number $\overline{\alpha}_e$ can be obtained by processing the derivation of robust objective function with respect to the design variables $x_e$ as

$$\overline{\alpha}_e = -\frac{1}{p}\frac{\partial \overline{C}}{\partial x_e} = -\frac{1}{p}\left(\frac{\partial \overline{E}(C)}{\partial x_e} + \kappa \frac{\partial \overline{SD}(C)}{\partial x_e}\right) \tag{31}$$

where $x_e$ represents $x_a$ on macro scale and $x_i$ on micro scale, respectively, in consideration of the two-scale structure. Due to the load is independent of the design variable, the derivatives of maximal expectation and standard variance with respect to the design variable can be derived from Eq. (27) as below

$$\frac{\partial \overline{E}(C)}{\partial x_e} = \mathbf{F}^{\mathrm{T}}\left(\frac{\partial \mathbf{U}_0}{\partial x_e} + \sum_{J=1}^{M}\sum_{K=1}^{N}\left(\frac{\partial \mathbf{U}_{1,J}}{\partial x_e}\Delta\mu\left(X_J\left(Y_J^K\right)\right)S_{JK}^1 + \mathbf{U}_{1,J}\Delta\mu\left(X_J\left(Y_J^K\right)\right)\frac{\partial S_{1,JK}}{\partial x_e}\right)\right) \tag{32a}$$

$$\begin{aligned}
\frac{\partial \overline{SD}(C)}{\partial x_e} = \mathbf{F}^{\mathrm{T}}\Bigg\{\sum_{J=1}^{M}\Bigg\{&\left(\frac{\partial \mathbf{U}_{2,J}}{\partial x_e}\sigma\left(X_J\left(Y_J^{\mathrm{m}}\right)\right)S_{2,J} + \mathbf{U}_{2,J}\sigma\left(X_J\left(Y_J^{\mathrm{m}}\right)\right)\frac{\partial S_{2,J}}{\partial x_e}\right) \\
&+\sum_{K=1}^{N}\Bigg[\left(\frac{\partial \mathbf{U}_{3,JK}}{\partial x_e}\sigma\left(X_J\left(Y_J^{\mathrm{m}}\right)\right)\Delta\mu\left(X_J\left(Y_J^K\right)\right) + \frac{\partial \mathbf{U}_{2,J}}{\partial x_e}\Delta\sigma\left(X_J\left(Y_J^K\right)\right)\right)S_{3,JK} \\
&+\left(\mathbf{U}_{3,JK}\sigma\left(X_J\left(Y_J^{\mathrm{m}}\right)\right)\Delta\mu\left(X_J\left(Y_J^K\right)\right) + \mathbf{U}_{2,J}\Delta\sigma\left(X_J\left(Y_J^K\right)\right)\right)\frac{\partial S_{3,JK}}{\partial x_e}\Bigg]\Bigg\}\Bigg\}
\end{aligned} \tag{32b}$$

where $\dfrac{\partial \mathbf{U}_0}{\partial x_e}$, $\dfrac{\partial \mathbf{U}_{1,J}}{\partial x_e}$, $\dfrac{\partial \mathbf{U}_{2,J}}{\partial x_e}$, $\dfrac{\partial \mathbf{U}_{3,JK}}{\partial x_e}$ on both macro- and micro- scales can be derived from Eq. (26). For simplicity, the detailed expressions are given in the Appendix.

It is noted that the sign parts $S_{1,JK}$, $S_{2,J}$ and $S_{3,JK}$ are not continuous, thus the Heaviside projection method can be used to smooth the sign function [65, 66] as follows:

$$S\left(f\left(x_e\right)\right) = \tanh\left(\beta f\left(x_e\right)\right) \tag{33}$$

where $\beta$ is the coefficient. The partial derivatives of $S\left(f\left(x_e\right)\right)$ with respect to design variable can be obtained as follow

$$\frac{\partial S\left(f\left(x_e\right)\right)}{\partial x_e} = \left(1 - \tanh^2\left(\beta f\left(x_e\right)\right)\right)\beta \frac{\partial f\left(x_e\right)}{\partial x_e} \tag{34}$$

By employing Eq. (34), $\dfrac{\partial S_{1,JK}}{\partial x_e}$, $\dfrac{\partial S_{2,J}}{\partial x_e}$ and $\dfrac{\partial S_{3,JK}}{\partial x_e}$ in Eq. (32) can be derived.

*4.3  Concurrent design*

In most of the existing multi-scale topology optimization, the volume fractions on each scale were arbitrarily appointed separately. This artificial volume distribution limits the final structural performance. In the design considering uncertainty, the impact of such setting will be further amplified. Due to the advantages of discretization, BESO method can realize iteration under the same constraint by normalizing and sequencing the sensitivity of multiple scales. Thus, the macro and micro structures can be designed concurrently with a uniform weight constraint [14]. The normalization of the sensitivity can be carried out as follows

$$\xi_a = \overline{\alpha}_a \Big/ \frac{\partial m}{\partial x_a}, \tag{35a}$$

$$\xi_i = \overline{\alpha}_i \Big/ \frac{\partial m}{\partial x_i}, \tag{35b}$$

where $\xi_a$ and $\xi_i$ denote the normalized elemental sensitivity number at the macro-scale and micro-scale, respectively. $\dfrac{\partial m}{\partial x_a}$ and $\dfrac{\partial m}{\partial x_i}$ are the variations of the total weight $m$ with respect to the design variables at macro- and micro- scales, which can be separately derived from Eqs. (4) and (6a) as

$$\frac{\partial m}{\partial x_a} = V_a \rho^H \left( x_i \right) \tag{36a}$$

$$\frac{\partial m}{\partial x_i} = \sum_{a=1}^{NE} x_a V_a \frac{\partial \rho^H \left( x_i \right)}{\partial x_i} = \frac{V_i}{|Y|} \left( \rho_1 - \rho_2 \right) \sum_{a=1}^{NE} x_a V_a \tag{36b}$$

*4.4  Numerical implementation*

The flowchart of the proposed RCTO is shown in Fig. 4, and its detailed explanation is outlined as follows:

**Step 1:** <u>Initializing:</u> Carry out the finite element mesh. Initialize the original design of the macrostructure and the microstructure by defining $x_a$ and $x_i$.

**Step 2:** <u>BESO definition:</u> Define the BESO parameters such as the target weight fraction $W_f^*$, the evolutionary ratio $ER$ and the filter radius for macro-scale $r_{\min}^{\mathrm{mac}}$ and micro-scale $r_{\min}^{\mathrm{mic}}$.

**Step 3:** <u>Uncertainty dealing and inputting:</u> Using the hybrid interval random model shown in Eq. (17) model the uncertain parameters that has imprecise probability. Input some important distribution parameters, for instance, expectation and standard variance.

**Step 4:** <u>Homogenization:</u> Calculate the effective property of PUC by Eq. (6). Meanwhile, derive the partial derivatives of the effective elastic matrix and the effective density matrix with respect to the related uncertain parameter.

**Step 5:** <u>IHPA processing:</u> Perform IHPA as Section 3.3 shows. Carry out the maximal expectation $\overline{E}(C)$ and standard variance $\overline{SD}(C)$ of the robust objective function.

**Step 6:** <u>Sensitivity deriving, normalizing and filtering:</u> Calculate the sensitivity of the objective function to design variables at each scale as Section 4.2 and Appendix showing; Normalize the sensitivity by Eq. (35) to for the concurrent design; Filtering the normalized sensitivity to avoid numerical instabilities [5] as follow

$$\zeta_e = \frac{\sum_{e=1}^{L} w(r_{ne})\zeta_e}{\sum_{e=1}^{L} w(r_{ne})} \tag{37}$$

where $L$ denotes the total number of nodes in sub-domain $\Omega_e$. The sub-domain $\Omega_e$ is generated by drawing a circle of radius $r_{\min}$, and the $e$-th element is the center of the circle. $r_{ne}$ represents the distance between the center of element $e$ and element $ne$. $\zeta_e$ is the sensitivity number of the $e$-th element no matter it is at macro or micro scale. $w(r_{ne})$ is the linear weight factor defined by

$$w(r_{ne}) = \begin{cases} r_{\min} - r_{ne} & \text{for } r_{ne} < r_{\min} \\ 0 & \text{for } r_{ne} \geq r_{\min} \end{cases} \tag{38}$$

**Step 7:** <u>Optimization process stabilizing:</u> For the $k$-th iteration ($k > 1$), average the sensitivity with its history value as below

$$\xi_e = \frac{\zeta_e^k + \zeta_e^{k-1}}{2} \tag{39}$$

**Step 8:** <u>Multi-scale structure concurrently updating:</u> Reconstruct the macrostructure and composite material according to the ranking of the elemental sensitivity numbers at both scales. With the limitation of weight fraction $W_f^k$, the design variables of the element of high sensitivity are assigned to 1, the others are assigned to $10^{-6}$. As the result, the topologies of both scales are updated concurrently.

**Step 9:** <u>Weight fraction checking:</u> Repeat Steps 4-8 when the weight fraction of current iteration does not meet the target weight fraction. And then determine the target weight fraction of the two-scale system for the next iteration as follows

$$W_f^{k+1} = W_f^k \left(1 - ER\right), \tag{40}$$

$$W_f^{k+1} = W_f^k \left(1 + ER\right). \tag{41}$$

When the current weight fraction $W_f^k$ is larger than $W_f^*$, reduce the weight fraction as Eq. (40); otherwise increase the weight fraction as Eq. (41). If the resulting $W_f^{k+1}$ is larger than $W_f^k$, then $W_f^{k+1}$ is set to $W_f^*$.

**Step 10:** <u>Convergence checking:</u> Repeat Step 4-9 until the objective function is convergent. The convergence check is expressed as follows:

$$error = \frac{\left| \sum_{\gamma=1}^{N} C_{k-\gamma+1} - \sum_{\gamma=1}^{N} C_{k-N-\gamma+1} \right|}{\sum_{\gamma=1}^{N} C_{k-\gamma+1}} \leq \tau \tag{42}$$

where $C_k$ represents the objective function value of the $k$-th iteration. $N$ is set to be 5, which means that the change of the objective function in the last 10 iterations is small enough. $\tau$ denotes the tolerance of change.

**Step 11:** <u>End:</u> Output the final robust design of structure and its composite material.

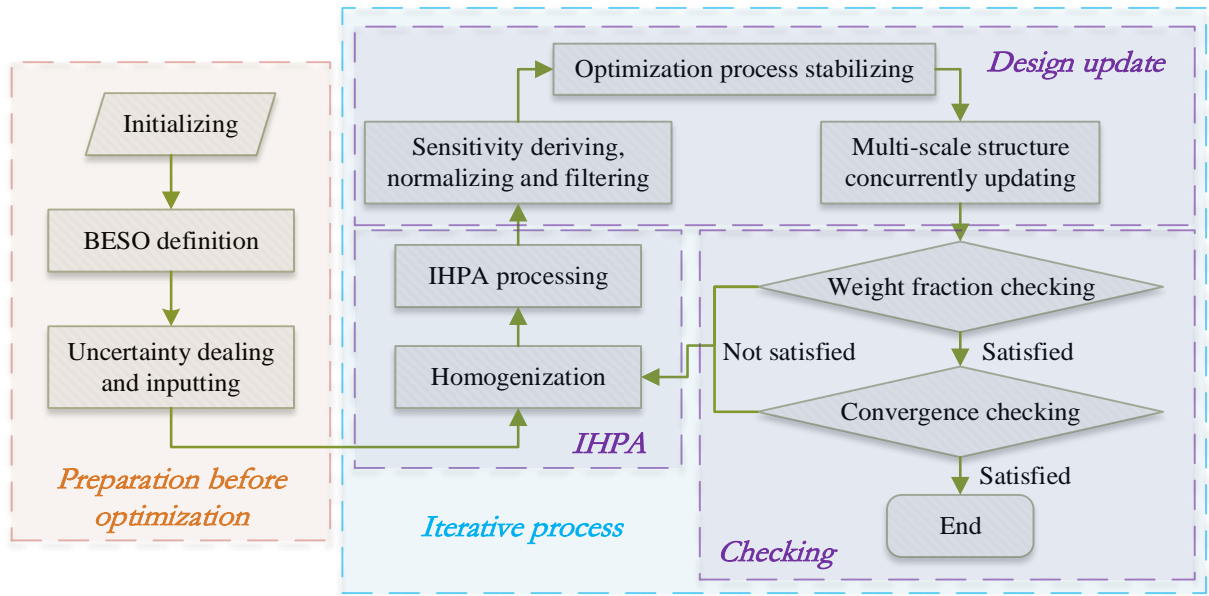

**Fig. 4** Flowchart of the RCTO procedure

## 5. Numerical examples

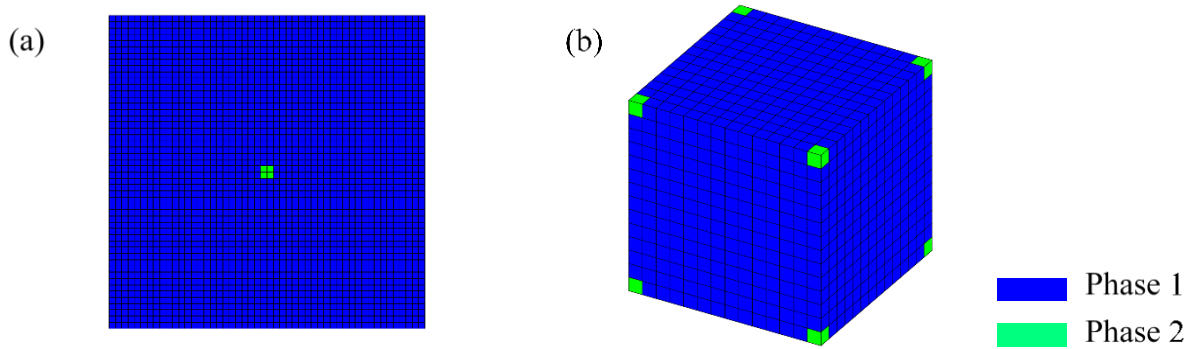

**Fig. 5** Initial designs of PUC: (a) 2D case; (b) 3D case.

In this section, both 2D and 3D examples are presented to prove the effectiveness of the proposed method. For 2D cases, the design domain is discretized by four nodes quadrilateral elements at both scales. The size of PUCs is $1\text{mm} \times 1\text{mm}$, which is divided into a $50 \times 50$ finite element mesh. As for 3D case, the design domain is meshed by hexahedral elements with $1\text{mm} \times 1\text{mm} \times 1\text{mm}$ size, which is divided into a $14 \times 14 \times 14$ finite element mesh. Fig. 5 shows the initial design of PUC, where the elements in blue denote phase 1 and the green ones represent phase 2. Based on the BESO framework, the initial design of macro-structure is a full design with initial design variable $x_a = 1$. $x_{\min} = 10^{-6}$ is adopted in this work. The penalty parameters $p$ at both scales are 3 and the evolutionary ratio $ER$ is 0.02. The filter radius is 3 times bigger than the elemental side length at each scale.

Table 1 shows the parameters of the probabilistic material properties, which is assumed to follow normal distribution, but the precise information is missing due to the difficulty and high cost of testing. Only the interval of their variation is obtained. It is noted that the Young's modulus and density are normally distributed, while normal distribution is not bounded. There might be some extreme conditions, where the Young's modulus or density has negative value. To solve such rare phenomena, readers can refer to [66].

**Table 1** Material properties in this study with imprecise probability.

| Material | Distribution parameter | Young's modulus (GPa) | Poisson's ratio | Density (g/mm³) |
|---|---|---|---|---|
| Phase 1 | Expectation | $[190,210]$ | $[0.285,0.315]$ | $[7900,8100]$ |
| | Standard variance | $[19,21]$ | $[0.001425,0.001575]$ | $[790,810]$ |
| Phase 2 | Expectation | $[140,160]$ | $[0.285,0.315]$ | $[790,810]$ |
| | Standard variance | $[14,16]$ | $[0.001425,0.001575]$ | $[79,81]$ |

*6.1 2D long cantilever beam with different robust optimization parameter $\kappa$*

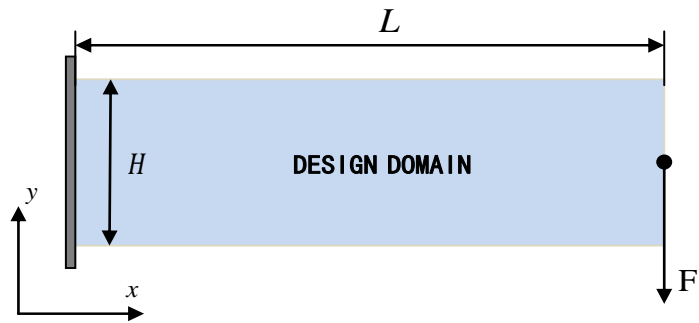

**Fig. 6** 2D long cantilever beam structure

In the first numerical example, various robust optimization parameters $\kappa$ are employed to prove the effect of the proposed RCTO method. The design domain, loading and boundary condition of the 2D cantilever beam are shown in Fig. 6. The excitation **F** is a periodic force with a magnitude of 1000N in 500Hz. The length and height of the design domain are 120mm and 40mm respectively. The target weight constraint is set to be 50%. Three different robust optimization weight parameters $\kappa=1$, $\kappa=3$ and $\kappa=5$ are employed.

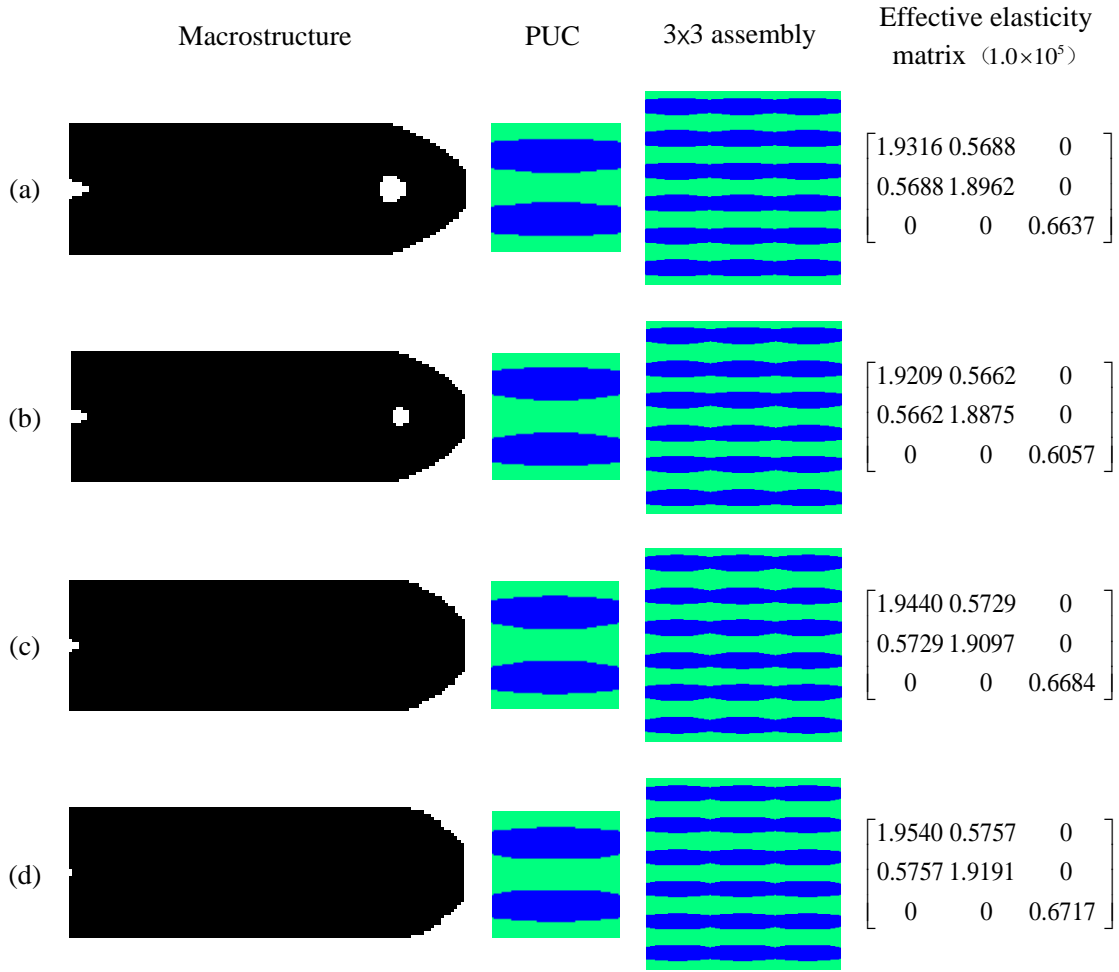

|  | Macrostructure | PUC | 3x3 assembly | Effective elasticity matrix $(1.0\times10^5)$ |

(a) $\begin{bmatrix} 1.9316 & 0.5688 & 0 \\ 0.5688 & 1.8962 & 0 \\ 0 & 0 & 0.6637 \end{bmatrix}$

(b) $\begin{bmatrix} 1.9209 & 0.5662 & 0 \\ 0.5662 & 1.8875 & 0 \\ 0 & 0 & 0.6057 \end{bmatrix}$

(c) $\begin{bmatrix} 1.9440 & 0.5729 & 0 \\ 0.5729 & 1.9097 & 0 \\ 0 & 0 & 0.6684 \end{bmatrix}$

(d) $\begin{bmatrix} 1.9540 & 0.5757 & 0 \\ 0.5757 & 1.9191 & 0 \\ 0 & 0 & 0.6717 \end{bmatrix}$

**Fig. 7** The macrostructure, PUC, 3x3 assembled PUC and the effective elasticity matrix designed by:(a) DCTO; (b) $\kappa=1$; (c) $\kappa=3$; (d) $\kappa=5$.

Fig. 7 shows the results of DCTO and RCTO, in which the FEA calls for each iteration of the two methods are 1 and 16, respectively. The topological designs of macrostructure and composite microstructure can be compared directly. By comparing the topological designs of DCTO and RCTO, it can be seen that there are differences in both layouts of macrostructure and PUC. By applying different robust parameter $\kappa$, which means changing the weighting distribution of expectation and standard variance, the final robust designs will be different. With the increase of robust parameters, the robust design is more inclined to adjust the macrostructure first and then the PUC. For the convenience of comparison, the $3 \times 3$ assembled PUCs and the effective elasticity matrix of each composite material are presented. These results show that RCTO can figure out topological designs different from DCTO, and in such working condition, the macrostructure is more sensitive to the uncertainty.

**Table 2** Value of the objective function of results shown in Fig. 7 considering uncertainties.

| Method | | Value of objective function | Difference |
|---|---|---|---|
| DCTO | | 842.3289 | |
| RCTO | $\kappa=1$ | 830.5541 | -11.7748 |
| | $\kappa=3$ | 824.9767 | -17.3522 |
| | $\kappa=5$ | 824.5547 | -17.7742 |

Table 2 lists the corresponding value of objective function obtained from the proposed IHPA method of the above optimization results with considering uncertainties. It can be seen that the mean-compliance of the RCTO designs are lower than the DCTO-based design, which means that the proposed RCTO performs better than the DCTO when it comes uncertainty. The value of mean-compliance decreases gradually with increasing $\kappa$, but the decrease tends to decrease: there is slightly difference between $\kappa=3$ and $\kappa=5$. It is because the objective function is weighed by the expectation and standard variance. Excessive standard variance weighting coefficient will reduce the influence of expectation on the result, but the value of expectation is much higher than the standard variance.

To verify the accuracy of the proposed IHPA method, the maximal value of expectation and standard variance are shown in Table 3 to compare with the results simulated by MCS. In the implementation of the MCS, the sample size is $10^6$, where the sample size of the random variable is $10^3$, and $10^3$ groups of interval values are involved in each random variable. The computational costs of the IHPA and MCS are also presented by the number of FEA calls. Two conclusions can be drawn from the comparison: On the one hand, there are some errors of the IHPA results compared to the MCS, which are within acceptable limits. On the other hand, the IHPA calculates the worst-case objective function with only 16 FEA calls, which is a great improvement over the MCS that requires $10^6$ times of FEA calls. The example shows that in the CTO field, the proposed IHPA-based RCTO method is an effective RCTO method with high calculation efficiency and only a little loss of accuracy.

**Table 3** Comparison of IHPA and MCS

| | Maximal value | IHPA | MCS | Relative errors |
|---|---|---|---|---|
| DCTO | Expectation | 764.8803 | 773.2563 | 1.08% |
| | Standard variance | 77.4486 | 82.3656 | 5.97% |
| | Objective function | 842.3289 | 855.6219 | 1.55% |
| RCTO $\kappa=1$ | Expectation | 756.0140 | 766.1026 | 1.31% |
| | Standard variance | 74.5400 | 78.6790 | 5.26% |
| | Objective function | 830.5541 | 844.7816 | 1.68% |
| $\kappa=3$ | Expectation | 750.7632 | 759.4720 | 1.16% |
| | Standard variance | 74.2135 | 78.6515 | 5.98% |
| | Objective function | 824.9767 | 838.1235 | 1.57% |
| $\kappa=5$ | Expectation | 750.4522 | 757.3564 | 0.91% |
| | Standard variance | 74.1025 | 79.3045 | 6.56% |
| | Objective function | 824.5547 | 836.6609 | 1.45% |
| FEA calls | | 16 | $10^6$ | |

*6.2 2D MBB beam with different weight constraint*

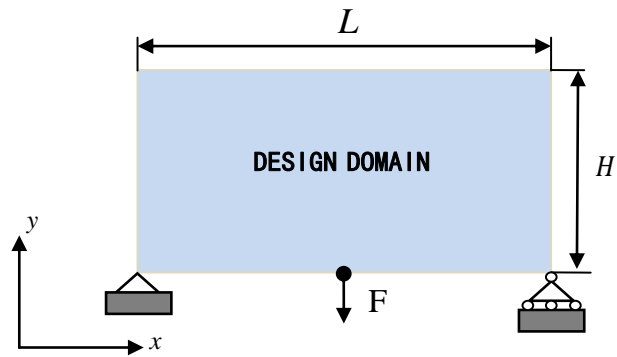

**Fig. 8** The design domain, boundary condition and loading of the MBB beam

In this example, we employed three different weight constraints: 75%, 40% and 5% to demonstrate the computational stability of the proposed RCTO method in dealing with imprecise probability. Fig. 8 shows that a force is applied at the bottom center of the 2D MBB beam. The magnitude of the force is 1000N and frequency is 2000Hz. The length of the beam is 90mm and the height is 40mm. The robust optimization parameter $\kappa$ is equal to 1.

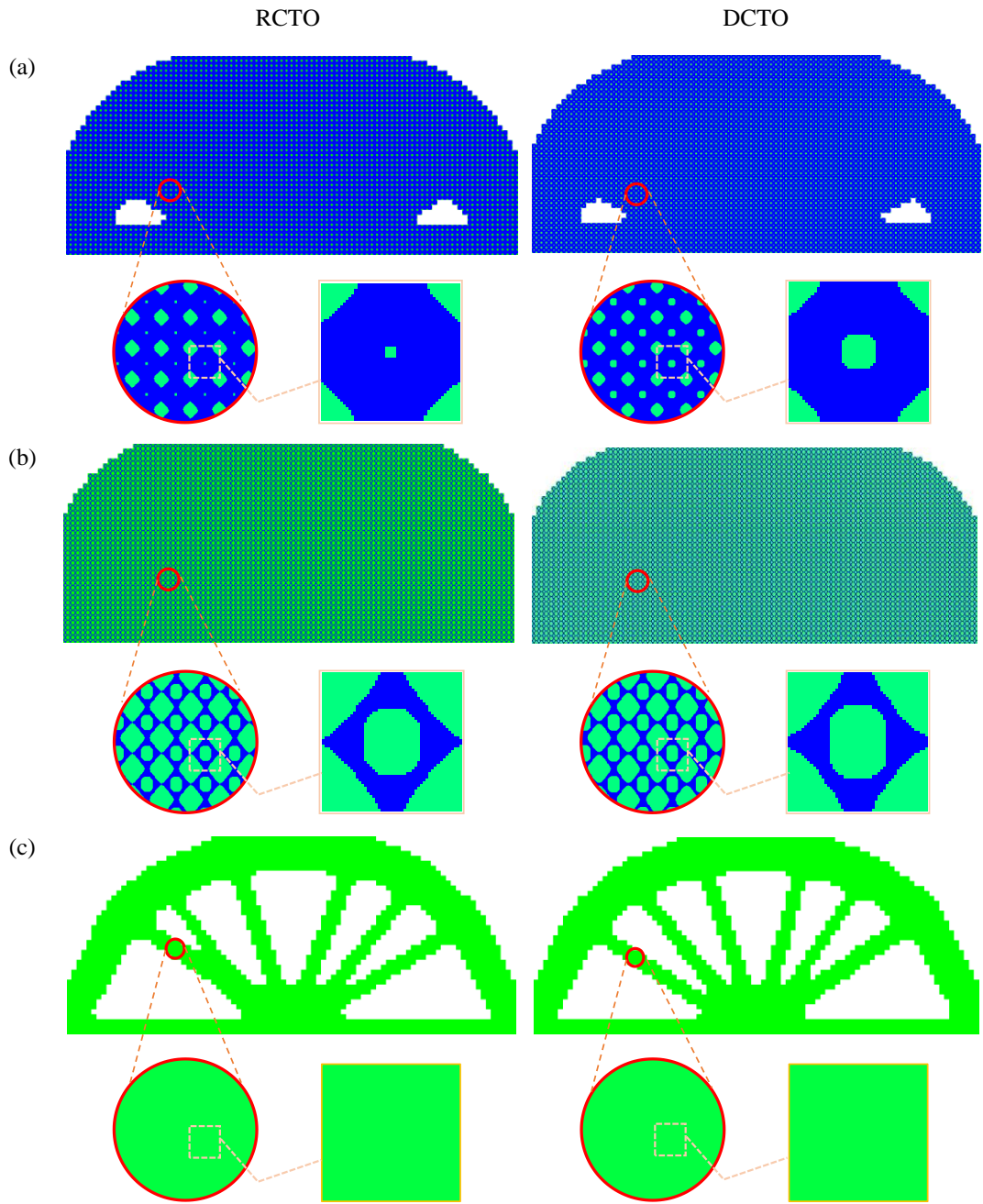

**Fig. 9** Topological designs of the 2D MBB beam acquired by RCTO (left) and DCTO (right) with different target weight fraction: (a) $W_f^* = 75\%$ ; (b) $W_f^* = 40\%$ ; (c) $W_f^* = 5\%$ .

Fig. 9 depicts the topological designs of RCTO and DCTO with different weight constraints.

The optimal distributions of macrostructure and PUC have different forms under different target weight fraction. By comparing the robust and deterministic design under different volumes, the topological distributions of the two designs are also significantly different. When $W_f^* = 75\%$, there are differences between macro and micro scales, but the difference between microstructure is more obvious. For $W_f^* = 40\%$, the difference between the two designs is mainly reflected in the microstructure. However, when $W_f^* = 5\%$, because of the total mass is too small, phase 2 has filled the microstructure, and the difference of the designs is totally on the macrostructure. For a more intuitive comparison, Table 4 lists the volume fractions of the solid element (at macro-scale) and phase 1 (at micro-scale) for the designs shown in Fig. 9.

Table 5 compares the maximal value of the objective function of these designs and all the values are calculated by IHPA. The comparison shows that under the same weight fraction, the result obtained by RCTO is always smaller than that obtained by DCTO. This proves that the proposed RCTO method performs better than DCTO in the dealing with uncertainty with imprecise probability.

Fig. 10 presents the iteration history of RCTO with different target weight fraction, where Fig. 10(a), Fig. 10(b) and Fig. 10(c) show the iterative curves of the structure under the weight fraction constraints of 75%, 40% and 5%, respectively. To represent the iterative process clearly, some intermediate topology designs in the iteration history are also demonstrated. It can be seen that the design of macrostructure and composite material are interactive with each other until the optimizations converge. It is worth mentioning that in Fig. 10(c), when the total mass is so small that phase 1 can no longer be tolerated, then the two-scale topology optimization has only been carried out at the macro scale. The subsequent iterative process is similar to the traditional single-scale topology optimization process, and the objective function increases as the structure decreases, eventually reaching convergence. In general, all iteration curves have clearly shown that the proposed method has a very good computational stability.

**Table 4** Volume fraction of the designs shown in Fig. 9.

| Total weight constraint | Volume fraction | RCTO | DCTO |
|---|---|---|---|
| 75% | Solid element | 89.3% | 90.3% |
| | Phase 1 | 82.2% | 81.1% |
| 40% | Solid element | 95.3% | 95.1% |
| | Phase 1 | 35.5% | 35.6% |
| 5% | Solid element | 50% | 50% |
| | Phase 1 | 0% | 0% |

**Table 5** The value of objective function of the designs in Fig. 9.

| Weight constraint | Value of objective function | |
|---|---|---|
| | RCTO | DCTO |
| 0.75 | 73.0651 | 74.9310 |
| 0.4 | 78.8372 | 79.6482 |
| 0.05 | 277.1851 | 282.0695 |

(a)

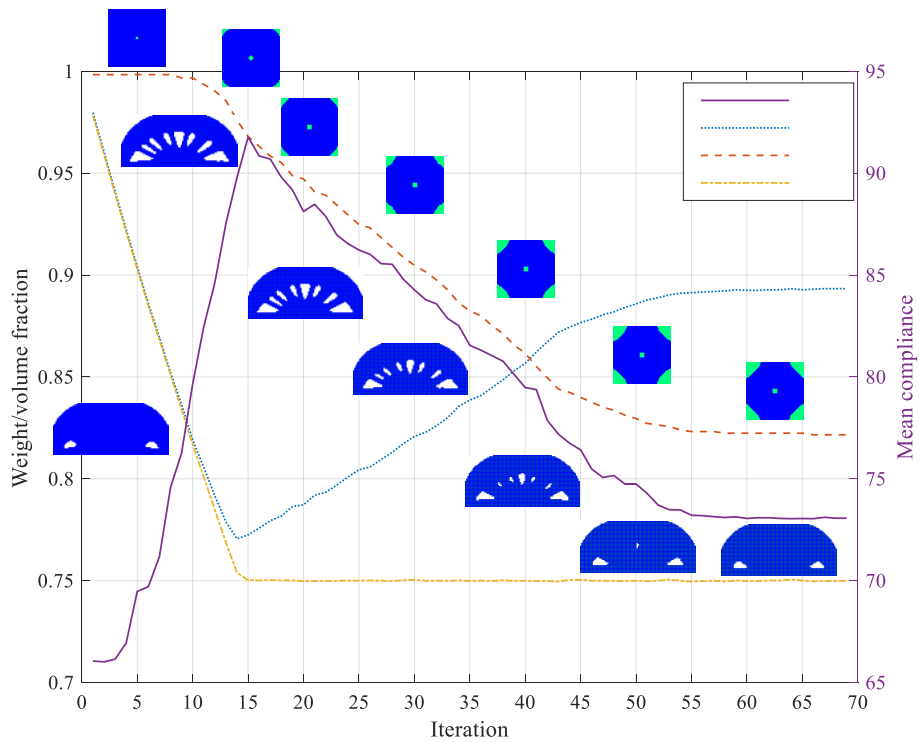

(b)

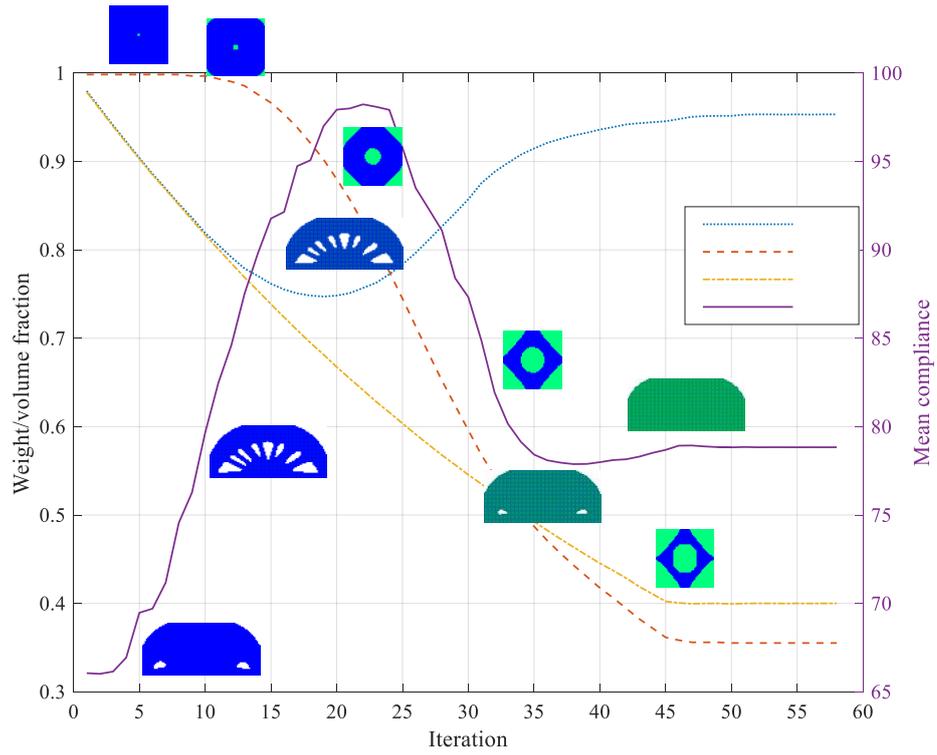

(c)

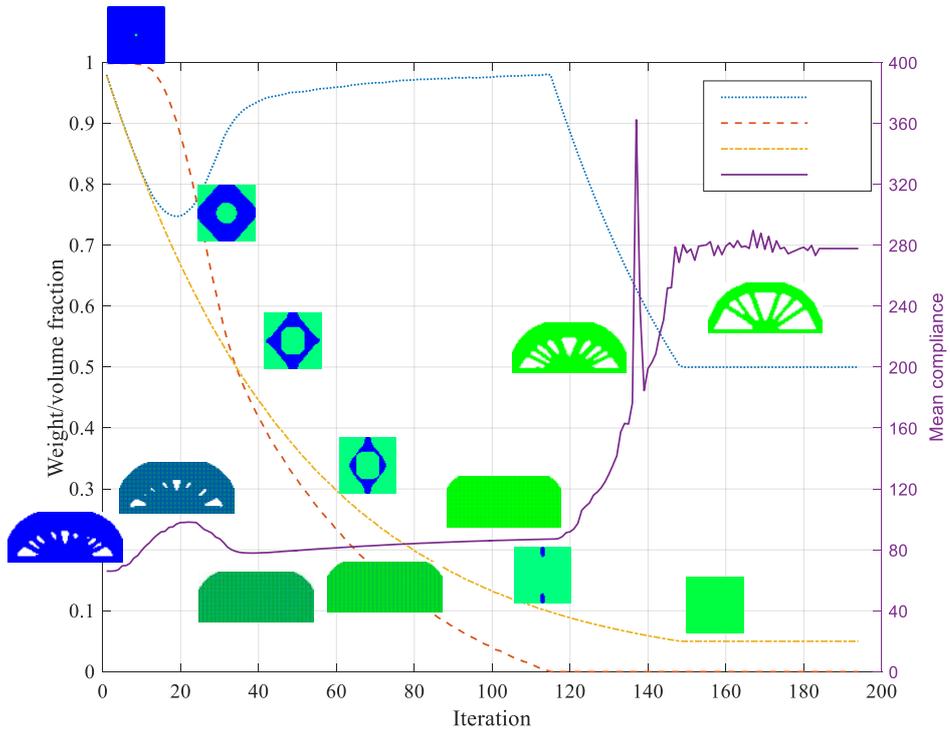

**Fig. 10** Iteration history of the 2D MBB beam with different target weight fraction: (a) $W_f^* = 75\%$ ;

(b) $W_f^* = 40\%$ ; (c) $W_f^* = 5\%$ .

*6.3 2D short cantilever beam in uniform/separate weight constraint*

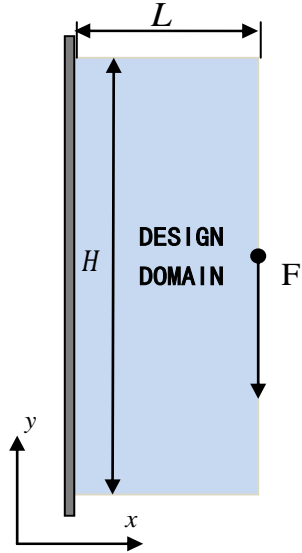

**Fig. 11** 2D short cantilever beam

In this example, we consider a 2D short cantilever that is loaded by multiple loading frequencies with an amplitude force of 1000N. Three different loading frequencies are selected to represent both of the static (0Hz) and dynamic (7500Hz and 15000Hz) loads. The boundary conditions and loading position are presented in Fig. 11. The length and height of the design domain are $L = 30\text{mm}$ and $H = 90\text{mm}$, respectively. and the robust optimization parameter is set as $\kappa = 1$. This example is presented to show the differences between the uniform weight constraint (UWC) and separate weight constraint (SWC), in which the SWC is guessed by 70% volume fractions on both of each scales so that the total weight constraint $W_f^*$ for both of the UWC and SWC can be 50%.

Fig. 12 shows the topological designs in different types of weight constraint, in which Fig. 12(a) shows the designs with UWC and Fig. 12(b) expresses the designs under SWC. It can be found that even if UWC and SWC use the same strategy to deal with uncertainty, there are many differences on the design of both macro and micro scales. For UWC, the volume fraction of the macrostructure increases as the frequency increases, while the volume fraction of phase 1 at the micro-scale is the opposite. Whereas for SWC, the volume fraction for both scales are constant, the design of its two scales can only seek changes under such constraints, which reduces the freedom of design and might lead to non-optimal design. To further understand the impact of the weight constraint setting, Table 6 lists the objective function value for the designs shown in Fig. 12. It is clear that at different frequencies, the results obtained through UWC are always lower than those obtained through SWC, which indicates that the UWC setting might be more suitable for RCTO to design structures with robust layout.

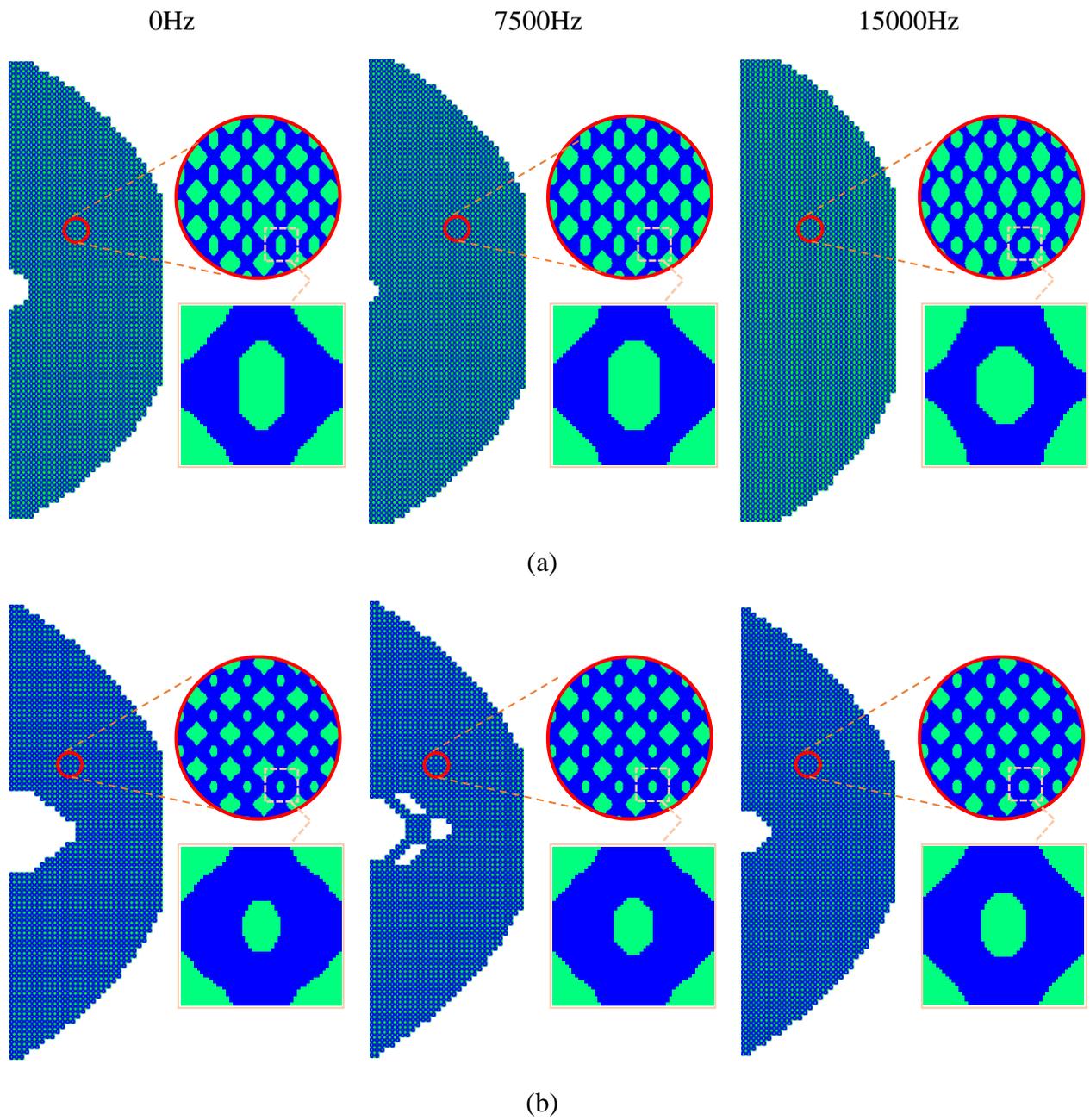

(a)

(b)

**Fig. 12** Topological designs of 2D short cantilever beam in different types of weight constraint with various loading frequencies: (a) UWC; (b) SWC.

**Table 6** The objective function of the topological designs shown in Fig. 12.

| Frequency (Hz) Strategy | 0 | 7500 | 15000 |
|---|---|---|---|
| UWC | 25.0745 | 25.1753 | 25.7462 |
| SWC | 27.0193 | 27.5028 | 28.6196 |

*6.4  3D prismatic structure in various loading frequency*

Fig. 13 depicts a 3D prismatic structure with fixed left side. The length, width and height of the structure are $L = 24\,\text{mm}$, $W = 8\,\text{mm}$ and $H = 8\,\text{mm}$, respectively. The design domain is discretized into $24 \times 8 \times 8$ eight-node hexahedral elements. A periodic force $\mathbf{F}$ with a magnitude of 1000N is loaded on the right bottom of the structure. The weight constraint is defined as 70% in this example. Three different excitation frequencies 0 Hz, 1000Hz and 2000Hz are considered for representing the static and dynamic conditions, respectively. The robust optimization weight parameter is set to $\kappa = 1$.

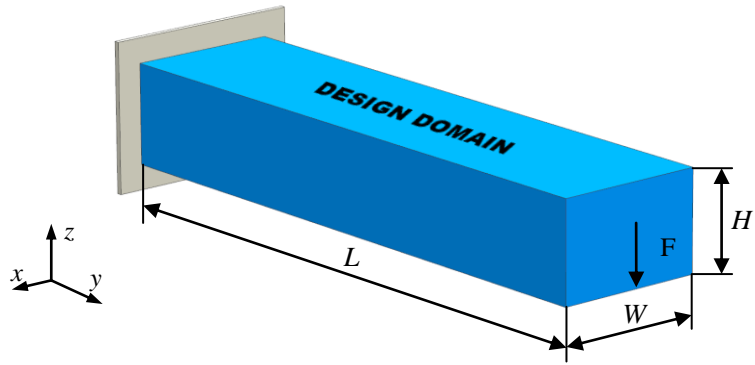

**Fig. 13** 3D cantilever

Fig. 14 shows the final designs of the two-scale 3D prismatic structure that is under 0Hz, 1000Hz and 2000Hz excitations. The volume fraction of solid and phase 1 on the corresponding scale are all shown in the table. The RCTO and DCTO designs can be compared intuitively. It can be observed that both approaches work well. The corresponding optimized topology of the macro-structures and composite microstructures are figured out. The results that the final designs of the structure are different due to the loading frequency, which indicates that it is necessary to optimize separately for different frequencies. By comparing the optimization results under different excitation frequencies, it can be found that the topology configurations obtained by RCTO and DCTO are also quite different for the existence of uncertainty. Such differences in design will lead to performance changes under uncertain conditions.

Table 7 shows the value of objective function of different topological designs in Fig. 14. All the results are calculated by the IHPA method, which has been approved to be efficient. By comparing the worst value of the mean-compliance, it is found that the results of RCTO are smaller than that of DCTO. That means that the proposed method can better realize concurrent topology optimization under imprecise uncertainty. This proves the robustness of the proposed RCTO method, which performs better than the traditional DCTO method under the hybrid interval random modeled imprecise uncertainty, again.

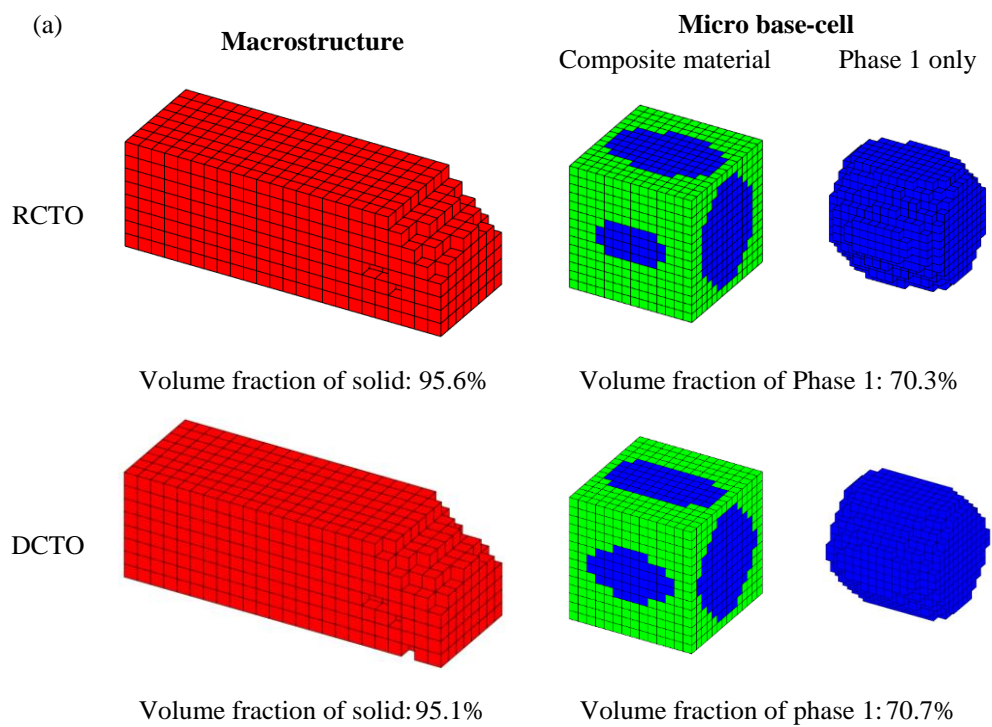

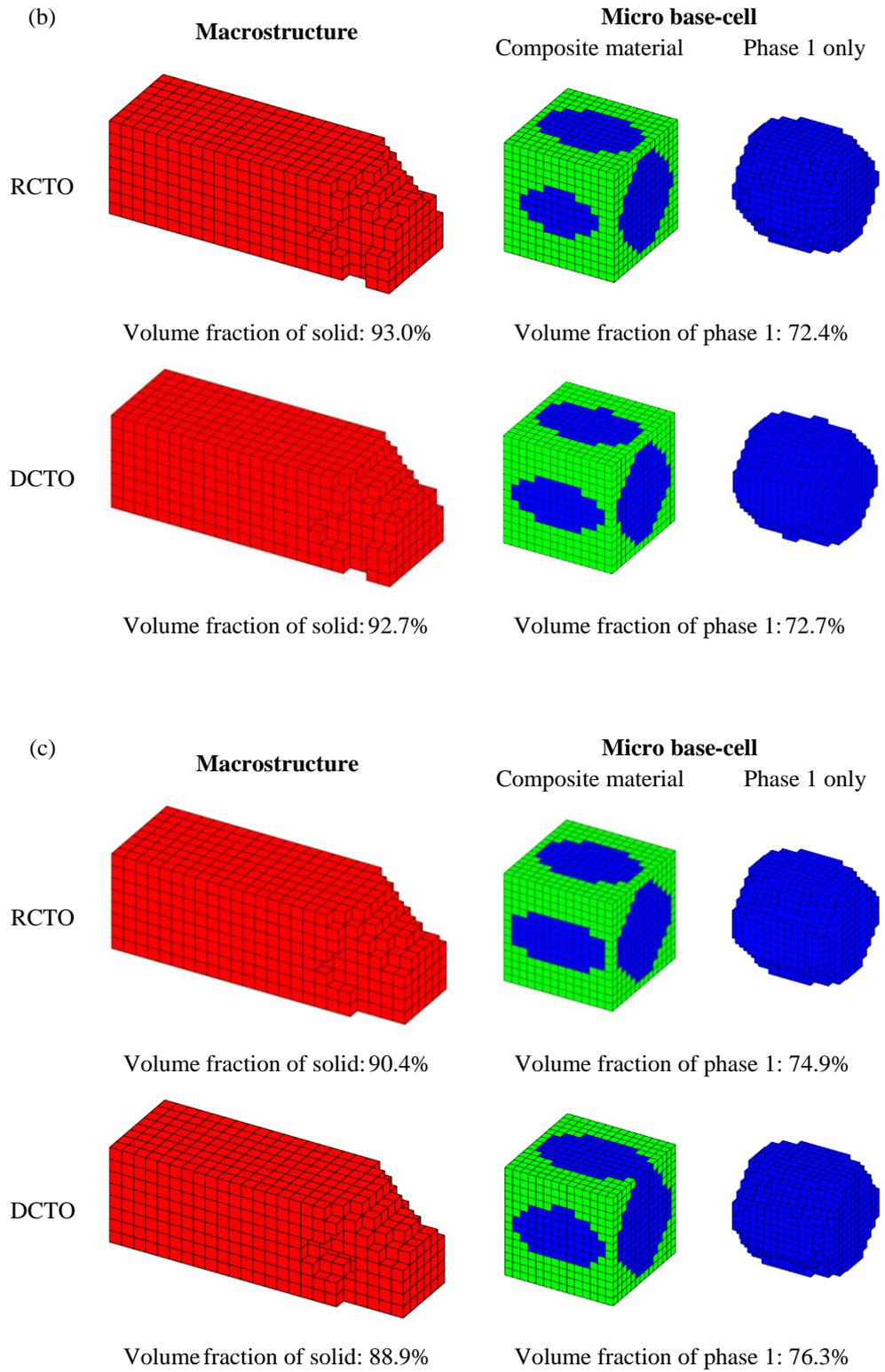

**Fig. 14** Topological designs of the two-scale 3D prismatic structure under different loading frequencies: (a) 0Hz; (b) 1000Hz; (c) 2000Hz.

**Table 7** The objective function values of the relative topological designs shown in Fig. 14.

| Frequency (Hz) | Value of objective function | |
| :---: | :---: | :---: |
| | RCTO | DCTO |
| 0 | 57.5272 | 57.8044 |
| 1000 | 59.6613 | 59.7918 |
| 2000 | 63.3583 | 64.1979 |

## 6. Conclusions

This work studied a robust concurrent topology optimization (RCTO) for the structure and its composite material considering imprecise uncertainty. For the first time, the material properties with imprecise probability are considered in the concurrent topology optimization. An improved hybrid perturbation analysis (IHPA) method is established to assess the worst performance of the structure under the type I hybrid interval random model based imprecise probability. By adopting IHPA, the RCTO approach for designing structure and its composite material is established. Both of 2D and 3D numerical examples are presented, in which the proposed RCTO method is proven to have high calculation efficiency, small precision loss and stable optimization process. Two types of weight constraint strategies are discussed, and it is found that the uniform weight constraint is more suitable for the robust designing in this work. Various excitation frequencies are taken into account, which shows that the proposed method performs well in both linear static and dynamic conditions. In general, the proposed RCTO has a great potential to be applied to robust multi-scale topology optimization involving uncertainty with imprecise probability, but may not be applicable to the problem of high nonlinearity and strong correlation uncertainties.

## Acknowledgements

This project is supported by the Foundation for Innovative Research Groups of the National Natural Science Foundation of China (Grant No. 51621004), and the Opening Project of Guangxi Key Laboratory of Automobile Components and Vehicle Technology, Guangxi University of Science and Technology (No. 2017GKLACVTKF01) and the Natural Science Foundation of Hunan Province, China (Grant No. 2017JJ3030). The authors would like to thank Dr. Jing Zheng (Hunan University) for her constructive suggestions on hybrid uncertainty.

## Appendix

The partial derivative of Eq. (26) with respect to $x_e$ can be derived as follows

$$\frac{\partial \mathbf{U}_0}{\partial x_e} = -\frac{\partial \mathbf{K}_d^{-1}\left(\mu\left(\mathbf{X}\left(\mathbf{Y}^m\right)\right)\right)}{\partial x_e}\mathbf{F} \tag{A.1}$$

$$\frac{\partial \mathbf{U}_{1,J}}{\partial x_e} = -\frac{\partial \mathbf{K}_d^{-1}\left(\mu\left(\mathbf{X}\left(\mathbf{Y}^m\right)\right)\right)}{\partial x_e}\frac{\partial \mathbf{K}_d\left(\mu\left(X_J\left(\mathbf{Y}_J\right)\right)\right)}{\partial \mu\left(X_J\left(\mathbf{Y}_J^1\right)\right)}\mathbf{U}_0$$

$$-\mathbf{K}_d^{-1}\left(\mu\left(\mathbf{X}\left(\mathbf{Y}^m\right)\right)\right)\left(\frac{\partial^2 \mathbf{K}_d\left(\mu\left(X_J\left(\mathbf{Y}_J\right)\right)\right)}{\partial \mu\left(X_J\left(\mathbf{Y}_J^1\right)\right)\partial x_e}\mathbf{U}_0 + \frac{\partial \mathbf{K}_d\left(\mu\left(X_J\left(\mathbf{Y}_J\right)\right)\right)}{\partial \mu\left(X_J\left(\mathbf{Y}_J^1\right)\right)}\frac{\partial \mathbf{K}_d^{-1}\left(\mu\left(\mathbf{X}\left(\mathbf{Y}^m\right)\right)\right)}{\partial x_e}\mathbf{F}\right) \tag{A.2}$$

$$\frac{\partial \mathbf{U}_{2,J}}{\partial x_e} = -\frac{\partial \mathbf{K}_d^{-1}\left(\mu\left(\mathbf{X}\left(\mathbf{Y}^m\right)\right)\right)}{\partial x_e}\frac{\partial \mathbf{K}_d\left(X_J\left(\mathbf{Y}_J\right)\right)}{\partial X_J\left(\mathbf{Y}_J\right)}\mathbf{U}_0$$

$$-\mathbf{K}_d^{-1}\left(\mu\left(\mathbf{X}\left(\mathbf{Y}^m\right)\right)\right)\left(\frac{\partial^2 \mathbf{K}_d\left(\mu\left(X_J\left(\mathbf{Y}_J\right)\right)\right)}{\partial X_J\left(\mathbf{Y}_J\right)\partial x_e}\mathbf{U}_0 + \frac{\partial \mathbf{K}_d\left(\mu\left(X_J\left(\mathbf{Y}_J\right)\right)\right)}{\partial X_J\left(\mathbf{Y}_J\right)}\frac{\partial \mathbf{K}_d^{-1}\left(\mu\left(\mathbf{X}\left(\mathbf{Y}^m\right)\right)\right)}{\partial x_e}\mathbf{F}\right) \tag{A.3}$$

$$\frac{\partial \mathbf{U}_{3,JK}}{\partial x_e} = -\frac{\partial \mathbf{K}_d^{-1}\left(\mu\left(\mathbf{X}\left(\mathbf{Y}^m\right)\right)\right)}{\partial x_e}\left(2\frac{\partial \mathbf{K}_d\left(\mu\left(X_J\left(\mathbf{Y}_J\right)\right)\right)}{\partial X_J\left(\mathbf{Y}_J\right)}\mathbf{U}_{2,J} + \frac{\partial^2 \mathbf{K}_d\left(\mu\left(X_J\left(\mathbf{Y}_J\right)\right)\right)}{\partial X_J\left(\mathbf{Y}_J\right)\partial X_J\left(\mathbf{Y}_J^1\right)}\mathbf{U}_0\right)$$

$$-\mathbf{K}_d^{-1}\left(\mu\left(\mathbf{X}\left(\mathbf{Y}^m\right)\right)\right)\left\{2\left[\frac{\partial^2 \mathbf{K}_d\left(\mu\left(X_J\left(\mathbf{Y}_J\right)\right)\right)}{\partial X_J\left(\mathbf{Y}_J\right)\partial x_e}\mathbf{U}_{2,J} - \frac{\partial \mathbf{K}_d\left(\mu\left(X_J\left(\mathbf{Y}_J\right)\right)\right)}{\partial X_J\left(\mathbf{Y}_J\right)}\right.\right.$$

$$\left(\frac{\partial \mathbf{K}_d^{-1}\left(\mu\left(\mathbf{X}\left(\mathbf{Y}^m\right)\right)\right)}{\partial x_e}\frac{\partial \mathbf{K}_d\left(X_J\left(\mathbf{Y}_J\right)\right)}{\partial X_J\left(\mathbf{Y}_J\right)}\mathbf{U}_0 + \mathbf{K}_d^{-1}\left(\mu\left(\mathbf{X}\left(\mathbf{Y}^m\right)\right)\right)\left(\frac{\partial^2 \mathbf{K}_d\left(\mu\left(X_J\left(\mathbf{Y}_J\right)\right)\right)}{\partial X_J\left(\mathbf{Y}_J\right)\partial x_e}\mathbf{U}_0\right.\right.$$

$$\left.\left.\left.+\frac{\partial \mathbf{K}_d\left(\mu\left(X_J\left(\mathbf{Y}_J\right)\right)\right)}{\partial X_J\left(\mathbf{Y}_J\right)}\frac{\partial \mathbf{K}_d^{-1}\left(\mu\left(\mathbf{X}\left(\mathbf{Y}^m\right)\right)\right)}{\partial x_e}\mathbf{F}\right)\right)\right] + \frac{\partial^3 \mathbf{K}_d\left(\mu\left(X_J\left(\mathbf{Y}_J\right)\right)\right)}{\partial X_J\left(\mathbf{Y}_J\right)\partial X_J\left(\mathbf{Y}_J^1\right)\partial x_e}\mathbf{U}_0 \tag{A.4}$$

$$\left.+\frac{\partial^2 \mathbf{K}_d\left(\mu\left(X_J\left(\mathbf{Y}_J\right)\right)\right)}{\partial X_J\left(\mathbf{Y}_J\right)\partial X_J\left(\mathbf{Y}_J^1\right)}\frac{\partial \mathbf{K}_d^{-1}\left(\mu\left(\mathbf{X}\left(\mathbf{Y}^m\right)\right)\right)}{\partial x_e}\mathbf{F}\right\}$$

where the partial derivation of the nonsingular matrix $\mathbf{K}_d\left(\mu\left(\mathbf{X}\left(\mathbf{Y}^m\right)\right)\right)$ with respect to $x_e$ can be derived by the following equation:

$$\mathbf{K}_d\left(\mu\left(\mathbf{X}\left(\mathbf{Y}^m\right)\right)\right)\mathbf{K}_d^{-1}\left(\mu\left(\mathbf{X}\left(\mathbf{Y}^m\right)\right)\right) = \mathbf{I} \tag{A.5}$$

Taking the partial derivation of both sides of Eq. (A.5) with respect to design variable $x_e$ yields

$$\frac{\partial \mathbf{K}_d^{-1}\left(\mu\left(\mathbf{X}\left(\mathbf{Y}^m\right)\right)\right)}{\partial x_e} = -\mathbf{K}_d^{-1}\left(\mu\left(\mathbf{X}\left(\mathbf{Y}^m\right)\right)\right)\frac{\partial \mathbf{K}_d\left(\mu\left(\mathbf{X}\left(\mathbf{Y}^m\right)\right)\right)}{\partial x_e}\mathbf{K}_d^{-1}\left(\mu\left(\mathbf{X}\left(\mathbf{Y}^m\right)\right)\right) \tag{A.6}$$

The equations $\dfrac{\partial \mathbf{K}_d\left(\mu\left(\mathbf{X}\left(\mathbf{Y}^m\right)\right)\right)}{\partial x_e}$, $\dfrac{\partial^2 \mathbf{K}_d\left(\mu\left(\mathbf{X}\left(\mathbf{Y}^m\right)\right)\right)}{\partial x_e \partial \mu\left(X_J\left(\mathbf{Y}_J^1\right)\right)}$, $\dfrac{\partial^2 \mathbf{K}_d\left(X_J\left(\mathbf{Y}_J\right)\right)}{\partial x_e \partial X_J\left(\mathbf{Y}_J\right)}$ and $\dfrac{\partial^3 \mathbf{K}_d\left(\mu\left(\mathbf{X}\left(\mathbf{Y}^m\right)\right)\right)}{\partial x_e \partial X_J\left(\mathbf{Y}_J\right) \partial X_J\left(\mathbf{Y}_J^1\right)}$ in

Eqs. (A.1)-(A.4) can be decomposed as

$$\frac{\partial \mathbf{K}_{d,e}\left(\mu\left(\mathbf{X}\left(\mathbf{Y}^m\right)\right)\right)}{\partial x_e} = \frac{\partial \mathbf{K}_e\left(\mu\left(\mathbf{X}\left(\mathbf{Y}^m\right)\right)\right)}{\partial x_e} - \omega^2 \frac{\partial \mathbf{M}_e\left(\mu\left(\mathbf{X}\left(\mathbf{Y}^m\right)\right)\right)}{\partial x_e}, \tag{A.7}$$

$$\frac{\partial^2 \mathbf{K}_{d,e}\left(\mu\left(\mathbf{X}\left(\mathbf{Y}^m\right)\right)\right)}{\partial x_e \partial \mu\left(X_J\left(\mathbf{Y}_J^1\right)\right)} = \frac{\partial^2 \mathbf{K}_e\left(\mu\left(\mathbf{X}\left(\mathbf{Y}^m\right)\right)\right)}{\partial x_e \partial \mu\left(X_J\left(\mathbf{Y}_J^1\right)\right)} - \omega^2 \frac{\partial^2 \mathbf{M}_e\left(\mu\left(\mathbf{X}\left(\mathbf{Y}^m\right)\right)\right)}{\partial x_e \partial \mu\left(X_J\left(\mathbf{Y}_J^1\right)\right)}, \tag{A.8}$$

$$\frac{\partial^2 \mathbf{K}_{d,e}\left(\mu\left(\mathbf{X}\left(\mathbf{Y}^m\right)\right)\right)}{\partial x_e \partial X_J\left(\mathbf{Y}_J\right)} = \frac{\partial^2 \mathbf{K}_e\left(\mu\left(\mathbf{X}\left(\mathbf{Y}^m\right)\right)\right)}{\partial x_e \partial X_J\left(\mathbf{Y}_J\right)} - \omega^2 \frac{\partial^2 \mathbf{M}_e\left(\mu\left(\mathbf{X}\left(\mathbf{Y}^m\right)\right)\right)}{\partial x_e \partial X_J\left(\mathbf{Y}_J\right)}, \tag{A.9}$$

$$\frac{\partial^3 \mathbf{K}_{d,e}\left(\mu\left(\mathbf{X}\left(\mathbf{Y}^m\right)\right)\right)}{\partial x_e \partial X_J\left(\mathbf{Y}_J\right) \partial X_J\left(\mathbf{Y}_J^1\right)} = \frac{\partial^3 \mathbf{K}_e\left(\mu\left(\mathbf{X}\left(\mathbf{Y}^m\right)\right)\right)}{\partial x_e \partial X_J\left(\mathbf{Y}_J\right) \partial X_J\left(\mathbf{Y}_J^1\right)} - \omega^2 \frac{\partial^3 \mathbf{M}_e\left(\mu\left(\mathbf{X}\left(\mathbf{Y}^m\right)\right)\right)}{\partial x_e \partial X_J\left(\mathbf{Y}_J\right) \partial X_J\left(\mathbf{Y}_J^1\right)}. \tag{A.10}$$

Note that $x_e$ stands for the design variable for both macro- and micro- scale.

### A.1 Sensitivity on macro-scale

On the macro-scale, $x_a$ is the design variable. Substituting Eq. (10) into Eqs. (A.7)-(A.10) yields the elemental sensitivity

$$\frac{\partial \mathbf{K}_a\left(\mu\left(\mathbf{X}\left(\mathbf{Y}^m\right)\right)\right)}{\partial x_a} = \begin{cases} p\int_A \mathbf{B}^T \mathbf{D}^H\left(\mu\left(\mathbf{X}\left(\mathbf{Y}^m\right)\right)\right)\mathbf{B}dA, & \text{when } x = 1 \\ px_{\min}^{p-1}\int_A \mathbf{B}^T \mathbf{D}^H\left(\mu\left(\mathbf{X}\left(\mathbf{Y}^m\right)\right)\right)\mathbf{B}dA, & \text{when } x = x_{\min} \end{cases} \tag{A.11}$$

$$\frac{\partial^2 \mathbf{K}_a\left(\mu\left(\mathbf{X}\left(\mathbf{Y}^m\right)\right)\right)}{\partial x_a \partial \mu\left(X_J\left(\mathbf{Y}_J^1\right)\right)} = \begin{cases} p\int_A \mathbf{B}^T \dfrac{\partial \mathbf{D}^H\left(\mu\left(\mathbf{X}\left(\mathbf{Y}^m\right)\right)\right)}{\partial \mu\left(X_J\left(\mathbf{Y}_J^1\right)\right)}\mathbf{B}dA, & \text{when } x = 1 \\[4mm] px_{\min}^{p-1}\int_A \mathbf{B}^T \dfrac{\partial \mathbf{D}^H\left(\mu\left(\mathbf{X}\left(\mathbf{Y}^m\right)\right)\right)}{\partial \mu\left(X_J\left(\mathbf{Y}_J^1\right)\right)}\mathbf{B}dA, & \text{when } x = x_{\min} \end{cases} \tag{A.12}$$

$$\frac{\partial^2 \mathbf{K}_a\left(\mu\left(\mathbf{X}\left(\mathbf{Y}^{\mathrm{m}}\right)\right)\right)}{\partial x_a \partial X_J\left(\mathbf{Y}\right)} = \begin{cases} p\int_A \mathbf{B}^T \dfrac{\partial \mathbf{D}^H\left(\mu\left(\mathbf{X}\left(\mathbf{Y}^{\mathrm{m}}\right)\right)\right)}{\partial X_J\left(\mathbf{Y}\right)} \mathbf{B} dA, & \text{when } x=1 \\[4mm] px_{\min}^{p-1}\int_A \mathbf{B}^T \dfrac{\partial \mathbf{D}^H\left(\mu\left(\mathbf{X}\left(\mathbf{Y}^{\mathrm{m}}\right)\right)\right)}{\partial X_J\left(\mathbf{Y}\right)} \mathbf{B} dA, & \text{when } x=x_{\min} \end{cases} \tag{A.13}$$

$$\frac{\partial^3 \mathbf{K}_a\left(\mu\left(\mathbf{X}\left(\mathbf{Y}^{\mathrm{m}}\right)\right)\right)}{\partial x_a \partial X_J\left(\mathbf{Y}_J\right)\partial X_J\left(\mathbf{Y}_J^1\right)} = \begin{cases} p\int_A \mathbf{B}^T \dfrac{\partial^2 \mathbf{D}^H\left(\mu\left(\mathbf{X}\left(\mathbf{Y}^{\mathrm{m}}\right)\right)\right)}{\partial X_J\left(\mathbf{Y}_J\right)\partial X_J\left(\mathbf{Y}_J^1\right)} \mathbf{B} dA, & \text{when } x=1 \\[4mm] px_{\min}^{p-1}\int_A \mathbf{B}^T \dfrac{\partial^2 \mathbf{D}^H\left(\mu\left(\mathbf{X}\left(\mathbf{Y}^{\mathrm{m}}\right)\right)\right)}{\partial X_J\left(\mathbf{Y}_J\right)\partial X_J\left(\mathbf{Y}_J^1\right)} \mathbf{B} dA, & \text{when } x=x_{\min} \end{cases} \tag{A.14}$$

$$\frac{\partial \mathbf{M}_a\left(\mu\left(\mathbf{X}\left(\mathbf{Y}^{\mathrm{m}}\right)\right)\right)}{\partial x_a} = \int_A \rho^H\left(\mu\left(\mathbf{X}\left(\mathbf{Y}^{\mathrm{m}}\right)\right)\right) \mathbf{N}^{\mathrm{T}} \mathbf{N} dA, \tag{A.15}$$

$$\frac{\partial^2 \mathbf{M}_a\left(\mu\left(\mathbf{X}\left(\mathbf{Y}^{\mathrm{m}}\right)\right)\right)}{\partial x_e \partial \mu\left(X_J\left(\mathbf{Y}_J^1\right)\right)} = \int_A \frac{\partial \rho^H\left(\mu\left(\mathbf{X}\left(\mathbf{Y}^{\mathrm{m}}\right)\right)\right)}{\partial \mu\left(X_J\left(\mathbf{Y}_J^1\right)\right)} \mathbf{N}^{\mathrm{T}} \mathbf{N} dA \tag{A.16}$$

$$\frac{\partial^2 \mathbf{M}_a\left(\mu\left(\mathbf{X}\left(\mathbf{Y}^{\mathrm{m}}\right)\right)\right)}{\partial x_e \partial X_J\left(\mathbf{Y}\right)} = \int_A \frac{\partial \rho^H\left(\mu\left(\mathbf{X}\left(\mathbf{Y}^{\mathrm{m}}\right)\right)\right)}{\partial X_J\left(\mathbf{Y}\right)} \mathbf{N}^{\mathrm{T}} \mathbf{N} dA, \tag{A.17}$$

$$\frac{\partial^3 \mathbf{M}_a\left(\mu\left(\mathbf{X}\left(\mathbf{Y}^{\mathrm{m}}\right)\right)\right)}{\partial x_e \partial X_J\left(\mathbf{Y}_J\right)\partial X_J\left(\mathbf{Y}_J^1\right)} = \int_A \frac{\partial^2 \rho^H\left(\mu\left(\mathbf{X}\left(\mathbf{Y}^{\mathrm{m}}\right)\right)\right)}{\partial X_J\left(\mathbf{Y}_J\right)\partial X_J\left(\mathbf{Y}_J^1\right)} \mathbf{N}^{\mathrm{T}} \mathbf{N} dA \tag{A.18}$$

### A.2 Sensitivity on micro-scale

On the micro-scale, $x_i$ is the design variable. Similarly, substituting Eqs. (12) and (13) into Eqs. (A.7)-(A.10) yields the elemental sensitivity as below

$$\frac{\partial \mathbf{K}_a\left(\mu\left(\mathbf{X}\left(\mathbf{Y}^{\mathrm{m}}\right)\right)\right)}{\partial x_i} = \begin{cases} \dfrac{p}{|Y|}\int_A \mathbf{B}^{\mathrm{T}}\left(\int_Y (\boldsymbol{\varepsilon}_0-\boldsymbol{\varepsilon})^{\mathrm{T}}\left(\mathbf{D}_1\left(\mu\left(\mathbf{X}\left(\mathbf{Y}^{\mathrm{m}}\right)\right)\right)-\mathbf{D}_2\left(\mu\left(\mathbf{X}\left(\mathbf{Y}^{\mathrm{m}}\right)\right)\right)\right)(\boldsymbol{\varepsilon}_0-\boldsymbol{\varepsilon})dY\right)\mathbf{B} dA, & \text{when } x_a=1, x_i=1 \\[4mm] \dfrac{px_{\min}^{p-1}}{|Y|}\int_A \mathbf{B}^{\mathrm{T}}\left(\int_Y (\boldsymbol{\varepsilon}_0-\boldsymbol{\varepsilon})^{\mathrm{T}}\left(\mathbf{D}_1\left(\mu\left(\mathbf{X}\left(\mathbf{Y}^{\mathrm{m}}\right)\right)\right)-\mathbf{D}_2\left(\mu\left(\mathbf{X}\left(\mathbf{Y}^{\mathrm{m}}\right)\right)\right)\right)(\boldsymbol{\varepsilon}_0-\boldsymbol{\varepsilon})dY\right)\mathbf{B} dA, & \text{when } x_a=1, x_i=x_{\min} \\[4mm] \dfrac{px_{\min}}{|Y|}\int_A \mathbf{B}^{\mathrm{T}}\left(\int_Y (\boldsymbol{\varepsilon}_0-\boldsymbol{\varepsilon})^{\mathrm{T}}\left(\mathbf{D}_1\left(\mu\left(\mathbf{X}\left(\mathbf{Y}^{\mathrm{m}}\right)\right)\right)-\mathbf{D}_2\left(\mu\left(\mathbf{X}\left(\mathbf{Y}^{\mathrm{m}}\right)\right)\right)\right)(\boldsymbol{\varepsilon}_0-\boldsymbol{\varepsilon})dY\right)\mathbf{B} dA, & \text{when } x_a=x_{\min}, x_i=1 \\[4mm] \dfrac{px_{\min}^{p}}{|Y|}\int_A \mathbf{B}^{\mathrm{T}}\left(\int_Y (\boldsymbol{\varepsilon}_0-\boldsymbol{\varepsilon})^{\mathrm{T}}\left(\mathbf{D}_1\left(\mu\left(\mathbf{X}\left(\mathbf{Y}^{\mathrm{m}}\right)\right)\right)-\mathbf{D}_2\left(\mu\left(\mathbf{X}\left(\mathbf{Y}^{\mathrm{m}}\right)\right)\right)\right)(\boldsymbol{\varepsilon}_0-\boldsymbol{\varepsilon})dY\right)\mathbf{B} dA, & \text{when } x_a=x_{\min}, x_i=x_{\min} \end{cases}$$

$$\tag{A.19}$$

$$\frac{\partial^2 \mathbf{K}_a\left(\mu\left(\mathbf{X}\left(\mathbf{Y}^{\mathrm{m}}\right)\right)\right)}{\partial x_i \partial \mu\left(X_J\left(\mathbf{Y}_J^{\mathrm{I}}\right)\right)} = \begin{cases} \dfrac{p}{|Y|}\int_A \mathbf{B}^{\mathrm{T}}\left(\int_Y (\boldsymbol{\varepsilon}_0-\boldsymbol{\varepsilon})^{\mathrm{T}}\dfrac{\partial\left(\mathbf{D}_1\left(\mu\left(\mathbf{X}\left(\mathbf{Y}^{\mathrm{m}}\right)\right)\right)-\mathbf{D}_2\left(\mu\left(\mathbf{X}\left(\mathbf{Y}^{\mathrm{m}}\right)\right)\right)\right)}{\partial\mu\left(X_J\left(\mathbf{Y}_J^{\mathrm{I}}\right)\right)}(\boldsymbol{\varepsilon}_0-\boldsymbol{\varepsilon})dY\right)\mathbf{B}dA, & \text{when } x_a=1, x_i=1 \\[2em] \dfrac{px_{\min}^{p-1}}{|Y|}\int_A \mathbf{B}^{\mathrm{T}}\left(\int_Y (\boldsymbol{\varepsilon}_0-\boldsymbol{\varepsilon})^{\mathrm{T}}\dfrac{\partial\left(\mathbf{D}_1\left(\mu\left(\mathbf{X}\left(\mathbf{Y}^{\mathrm{m}}\right)\right)\right)-\mathbf{D}_2\left(\mu\left(\mathbf{X}\left(\mathbf{Y}^{\mathrm{m}}\right)\right)\right)\right)}{\partial\mu\left(X_J\left(\mathbf{Y}_J^{\mathrm{I}}\right)\right)}(\boldsymbol{\varepsilon}_0-\boldsymbol{\varepsilon})dY\right)\mathbf{B}dA, & \text{when } x_a=1, x_i=x_{\min} \\[2em] \dfrac{px_{\min}}{|Y|}\int_A \mathbf{B}^{\mathrm{T}}\left(\int_Y (\boldsymbol{\varepsilon}_0-\boldsymbol{\varepsilon})^{\mathrm{T}}\dfrac{\partial\left(\mathbf{D}_1\left(\mu\left(\mathbf{X}\left(\mathbf{Y}^{\mathrm{m}}\right)\right)\right)-\mathbf{D}_2\left(\mu\left(\mathbf{X}\left(\mathbf{Y}^{\mathrm{m}}\right)\right)\right)\right)}{\partial\mu\left(X_J\left(\mathbf{Y}_J^{\mathrm{I}}\right)\right)}(\boldsymbol{\varepsilon}_0-\boldsymbol{\varepsilon})dY\right)\mathbf{B}dA, & \text{when } x_a=x_{\min}, x_i=1 \\[2em] \dfrac{px_{\min}^{p}}{|Y|}\int_A \mathbf{B}^{\mathrm{T}}\left(\int_Y (\boldsymbol{\varepsilon}_0-\boldsymbol{\varepsilon})^{\mathrm{T}}\dfrac{\partial\left(\mathbf{D}_1\left(\mu\left(\mathbf{X}\left(\mathbf{Y}^{\mathrm{m}}\right)\right)\right)-\mathbf{D}_2\left(\mu\left(\mathbf{X}\left(\mathbf{Y}^{\mathrm{m}}\right)\right)\right)\right)}{\partial\mu\left(X_J\left(\mathbf{Y}_J^{\mathrm{I}}\right)\right)}(\boldsymbol{\varepsilon}_0-\boldsymbol{\varepsilon})dY\right)\mathbf{B}dA, & \text{when } x_a=x_{\min}, x_i=x_{\min} \end{cases}$$

(A.20)

$$\frac{\partial^2 \mathbf{K}_a\left(\mu\left(\mathbf{X}\left(\mathbf{Y}^{\mathrm{m}}\right)\right)\right)}{\partial x_i \partial X_J(\mathbf{Y})} = \begin{cases} \dfrac{p}{|Y|}\int_A \mathbf{B}^{\mathrm{T}}\left(\int_Y (\boldsymbol{\varepsilon}_0-\boldsymbol{\varepsilon})^{\mathrm{T}}\dfrac{\partial\left(\mathbf{D}_1\left(\mu\left(\mathbf{X}\left(\mathbf{Y}^{\mathrm{m}}\right)\right)\right)-\mathbf{D}_2\left(\mu\left(\mathbf{X}\left(\mathbf{Y}^{\mathrm{m}}\right)\right)\right)\right)}{\partial X_J(\mathbf{Y})}(\boldsymbol{\varepsilon}_0-\boldsymbol{\varepsilon})dY\right)\mathbf{B}dA, & \text{when } x_a=1, x_i=1 \\[2em] \dfrac{px_{\min}^{p-1}}{|Y|}\int_A \mathbf{B}^{\mathrm{T}}\left(\int_Y (\boldsymbol{\varepsilon}_0-\boldsymbol{\varepsilon})^{\mathrm{T}}\dfrac{\partial\left(\mathbf{D}_1\left(\mu\left(\mathbf{X}\left(\mathbf{Y}^{\mathrm{m}}\right)\right)\right)-\mathbf{D}_2\left(\mu\left(\mathbf{X}\left(\mathbf{Y}^{\mathrm{m}}\right)\right)\right)\right)}{\partial X_J(\mathbf{Y})}(\boldsymbol{\varepsilon}_0-\boldsymbol{\varepsilon})dY\right)\mathbf{B}dA, & \text{when } x_a=1, x_i=x_{\min} \\[2em] \dfrac{px_{\min}}{|Y|}\int_A \mathbf{B}^{\mathrm{T}}\left(\int_Y (\boldsymbol{\varepsilon}_0-\boldsymbol{\varepsilon})^{\mathrm{T}}\dfrac{\partial\left(\mathbf{D}_1\left(\mu\left(\mathbf{X}\left(\mathbf{Y}^{\mathrm{m}}\right)\right)\right)-\mathbf{D}_2\left(\mu\left(\mathbf{X}\left(\mathbf{Y}^{\mathrm{m}}\right)\right)\right)\right)}{\partial X_J(\mathbf{Y})}(\boldsymbol{\varepsilon}_0-\boldsymbol{\varepsilon})dY\right)\mathbf{B}dA, & \text{when } x_a=x_{\min}, x_i=1 \\[2em] \dfrac{px_{\min}^{p}}{|Y|}\int_A \mathbf{B}^{\mathrm{T}}\left(\int_Y (\boldsymbol{\varepsilon}_0-\boldsymbol{\varepsilon})^{\mathrm{T}}\dfrac{\partial\left(\mathbf{D}_1\left(\mu\left(\mathbf{X}\left(\mathbf{Y}^{\mathrm{m}}\right)\right)\right)-\mathbf{D}_2\left(\mu\left(\mathbf{X}\left(\mathbf{Y}^{\mathrm{m}}\right)\right)\right)\right)}{\partial X_J(\mathbf{Y})}(\boldsymbol{\varepsilon}_0-\boldsymbol{\varepsilon})dY\right)\mathbf{B}dA, & \text{when } x_a=x_{\min}, x_i=x_{\min} \end{cases}$$

(A.21)

$$\frac{\partial^3 \mathbf{K}_a\left(\mu\left(\mathbf{X}\left(\mathbf{Y}^{\mathrm{m}}\right)\right)\right)}{\partial x_i \partial X_J(\mathbf{Y}_J)\partial X_J\left(\mathbf{Y}_J^{\mathrm{I}}\right)} = \begin{cases} \dfrac{p}{|Y|}\int_A \mathbf{B}^{\mathrm{T}}\left(\int_Y (\boldsymbol{\varepsilon}_0-\boldsymbol{\varepsilon})^{\mathrm{T}}\dfrac{\partial^2\left(\mathbf{D}_1\left(\mu\left(\mathbf{X}\left(\mathbf{Y}^{\mathrm{m}}\right)\right)\right)-\mathbf{D}_2\left(\mu\left(\mathbf{X}\left(\mathbf{Y}^{\mathrm{m}}\right)\right)\right)\right)}{\partial X_J(\mathbf{Y}_J)\partial X_J\left(\mathbf{Y}_J^{\mathrm{I}}\right)}(\boldsymbol{\varepsilon}_0-\boldsymbol{\varepsilon})dY\right)\mathbf{B}dA, & \text{when } x_a=1, x_i=1 \\[2em] \dfrac{px_{\min}^{p-1}}{|Y|}\int_A \mathbf{B}^{\mathrm{T}}\left(\int_Y (\boldsymbol{\varepsilon}_0-\boldsymbol{\varepsilon})^{\mathrm{T}}\dfrac{\partial^2\left(\mathbf{D}_1\left(\mu\left(\mathbf{X}\left(\mathbf{Y}^{\mathrm{m}}\right)\right)\right)-\mathbf{D}_2\left(\mu\left(\mathbf{X}\left(\mathbf{Y}^{\mathrm{m}}\right)\right)\right)\right)}{\partial X_J(\mathbf{Y}_J)\partial X_J\left(\mathbf{Y}_J^{\mathrm{I}}\right)}(\boldsymbol{\varepsilon}_0-\boldsymbol{\varepsilon})dY\right)\mathbf{B}dA, & \text{when } x_a=1, x_i=x_{\min} \\[2em] \dfrac{px_{\min}}{|Y|}\int_A \mathbf{B}^{\mathrm{T}}\left(\int_Y (\boldsymbol{\varepsilon}_0-\boldsymbol{\varepsilon})^{\mathrm{T}}\dfrac{\partial^2\left(\mathbf{D}_1\left(\mu\left(\mathbf{X}\left(\mathbf{Y}^{\mathrm{m}}\right)\right)\right)-\mathbf{D}_2\left(\mu\left(\mathbf{X}\left(\mathbf{Y}^{\mathrm{m}}\right)\right)\right)\right)}{\partial X_J(\mathbf{Y}_J)\partial X_J\left(\mathbf{Y}_J^{\mathrm{I}}\right)}(\boldsymbol{\varepsilon}_0-\boldsymbol{\varepsilon})dY\right)\mathbf{B}dA, & \text{when } x_a=x_{\min}, x_i=1 \\[2em] \dfrac{px_{\min}^{p}}{|Y|}\int_A \mathbf{B}^{\mathrm{T}}\left(\int_Y (\boldsymbol{\varepsilon}_0-\boldsymbol{\varepsilon})^{\mathrm{T}}\dfrac{\partial^2\left(\mathbf{D}_1\left(\mu\left(\mathbf{X}\left(\mathbf{Y}^{\mathrm{m}}\right)\right)\right)-\mathbf{D}_2\left(\mu\left(\mathbf{X}\left(\mathbf{Y}^{\mathrm{m}}\right)\right)\right)\right)}{\partial X_J(\mathbf{Y}_J)\partial X_J\left(\mathbf{Y}_J^{\mathrm{I}}\right)}(\boldsymbol{\varepsilon}_0-\boldsymbol{\varepsilon})dY\right)\mathbf{B}dA, & \text{when } x_a=x_{\min}, x_i=x_{\min} \end{cases}$$

$$\text{(A.22)}$$

$$\frac{\partial \mathbf{M}_a\big(\mu\big(\mathbf{X}\big(\mathbf{Y}^m\big)\big)\big)}{\partial x_i} = \begin{cases} \dfrac{1}{|Y|} \int_A \left( \displaystyle\sum_{i=1}^{Ne} V_i \Big( \rho_1\big(\mu\big(\mathbf{X}\big(\mathbf{Y}^m\big)\big)\big) - \rho_2\big(\mu\big(\mathbf{X}\big(\mathbf{Y}^m\big)\big)\big) \Big) \right) \mathbf{N}^{\mathrm{T}}\mathbf{N}dA, & \text{when } x_a = 1 \\[3ex] \dfrac{x_{\min}}{|Y|} \int_A \left( \displaystyle\sum_{i=1}^{Ne} V_i \Big( \rho_1\big(\mu\big(\mathbf{X}\big(\mathbf{Y}^m\big)\big)\big) - \rho_2\big(\mu\big(\mathbf{X}\big(\mathbf{Y}^m\big)\big)\big) \Big) \right) \mathbf{N}^{\mathrm{T}}\mathbf{N}dA, & \text{when } x_a = x_{\min} \end{cases}$$

$$\text{(A.23)}$$

$$\frac{\partial^2 \mathbf{M}_a\big(\mu\big(\mathbf{X}\big(\mathbf{Y}^m\big)\big)\big)}{\partial x_i \partial \mu\big(X_J\big(\mathbf{Y}_J^1\big)\big)} = \begin{cases} \dfrac{1}{|Y|} \int_A \left( \displaystyle\sum_{i=1}^{Ne} V_i \frac{\partial \Big( \rho_1\big(\mu\big(\mathbf{X}\big(\mathbf{Y}^m\big)\big)\big) - \rho_2\big(\mu\big(\mathbf{X}\big(\mathbf{Y}^m\big)\big)\big) \Big)}{\partial \mu\big(X_J\big(\mathbf{Y}_J^1\big)\big)} \right) \mathbf{N}^{\mathrm{T}}\mathbf{N}dA, & \text{when } x_a = 1 \\[3ex] \dfrac{x_{\min}}{|Y|} \int_A \left( \displaystyle\sum_{i=1}^{Ne} V_i \frac{\partial \Big( \rho_1\big(\mu\big(\mathbf{X}\big(\mathbf{Y}^m\big)\big)\big) - \rho_2\big(\mu\big(\mathbf{X}\big(\mathbf{Y}^m\big)\big)\big) \Big)}{\partial \mu\big(X_J\big(\mathbf{Y}_J^1\big)\big)} \right) \mathbf{N}^{\mathrm{T}}\mathbf{N}dA, & \text{when } x_a = x_{\min} \end{cases}$$

$$\text{(A.24)}$$

$$\frac{\partial^2 \mathbf{M}_a\big(\mu\big(\mathbf{X}\big(\mathbf{Y}^m\big)\big)\big)}{\partial x_i \partial X_J\big(\mathbf{Y}_J\big)} = \begin{cases} \dfrac{1}{|Y|} \int_A \left( \displaystyle\sum_{i=1}^{Ne} V_i \frac{\partial \Big( \rho_1\big(\mu\big(\mathbf{X}\big(\mathbf{Y}^m\big)\big)\big) - \rho_2\big(\mu\big(\mathbf{X}\big(\mathbf{Y}^m\big)\big)\big) \Big)}{\partial X_J\big(\mathbf{Y}_J\big)} \right) \mathbf{N}^{\mathrm{T}}\mathbf{N}dA, & \text{when } x_a = 1 \\[3ex] \dfrac{x_{\min}}{|Y|} \int_A \left( \displaystyle\sum_{i=1}^{Ne} V_i \frac{\partial \Big( \rho_1\big(\mu\big(\mathbf{X}\big(\mathbf{Y}^m\big)\big)\big) - \rho_2\big(\mu\big(\mathbf{X}\big(\mathbf{Y}^m\big)\big)\big) \Big)}{\partial X_J\big(\mathbf{Y}_J\big)} \right) \mathbf{N}^{\mathrm{T}}\mathbf{N}dA, & \text{when } x_a = x_{\min} \end{cases}$$

$$\text{(A.25)}$$

$$\frac{\partial^3 \mathbf{M}_a\big(\mu\big(\mathbf{X}\big(\mathbf{Y}^m\big)\big)\big)}{\partial x_i \partial X_J\big(\mathbf{Y}_J\big) \partial X_J\big(\mathbf{Y}_J^1\big)} = \begin{cases} \dfrac{1}{|Y|} \int_A \left( \displaystyle\sum_{i=1}^{Ne} V_i \frac{\partial^2 \Big( \rho_1\big(\mu\big(\mathbf{X}\big(\mathbf{Y}^m\big)\big)\big) - \rho_2\big(\mu\big(\mathbf{X}\big(\mathbf{Y}^m\big)\big)\big) \Big)}{\partial X_J\big(\mathbf{Y}_J\big) \partial X_J\big(\mathbf{Y}_J^1\big)} \right) \mathbf{N}^{\mathrm{T}}\mathbf{N}dA, & \text{when } x_a = 1 \\[3ex] \dfrac{x_{\min}}{|Y|} \int_A \left( \displaystyle\sum_{i=1}^{Ne} V_i \frac{\partial^2 \Big( \rho_1\big(\mu\big(\mathbf{X}\big(\mathbf{Y}^m\big)\big)\big) - \rho_2\big(\mu\big(\mathbf{X}\big(\mathbf{Y}^m\big)\big)\big) \Big)}{\partial X_J\big(\mathbf{Y}_J\big) \partial X_J\big(\mathbf{Y}_J^1\big)} \right) \mathbf{N}^{\mathrm{T}}\mathbf{N}dA, & \text{when } x_a = x_{\min} \end{cases}$$

$$\text{(A.26)}$$